\documentclass[prb,preprint,showpacs,showkeys,amsmath,amssymb,superscriptaddress,10pt]{revtex4}
\newcommand{\bra}[1]{\langle #1|}
\newcommand{\ket}[1]{|#1\rangle}
\newcommand{\braket}[2]{\langle #1|#2\rangle}
\usepackage{graphicx}
\usepackage{dcolumn}
\usepackage{bm}
\usepackage{graphicx}
\usepackage{epsfig}
\usepackage{amsmath}
\usepackage{amssymb}
\usepackage{natbib}
\usepackage{color}
\usepackage{tikz}
\usepackage{calc}
\usepackage{subfig}
\usepackage{frcursive}

\makeatletter
\newlength\@SizeOfCirc%
\newcommand{\CricArrowRight}[1]{%
    \setlength{\@SizeOfCirc}{\maxof{\widthof{#1}}{\heightof{#1}}}%
    \tikz [x=1.0ex,y=1.0ex,line width=.15ex, draw=black]%
        \draw [->,anchor=center]%
            node (0,0) {#1}%
            (0,1.2\@SizeOfCirc) arc (85:-240:1.2\@SizeOfCirc);%
}%
\makeatother

\begin{document}

\draft
\title{A bird's eye view on the flat and conic band world of the honeycomb and Kagome lattices: Towards an understanding of  2D Metal-Organic Frameworks  electronic structure.  }

\author{C. Barreteau}
\affiliation{SPEC, CEA, CNRS, Universit\'e Paris-Saclay, CEA Saclay 91191 Gif sur Yvette, France}

\author{F. Ducastelle}
\affiliation{Laboratoire d'Etude des Microstructures, ONERA-CNRS, BP 72, 92322 Ch\^{a}tillon Cedex, France}

\author{T. Mallah}
\affiliation{Institut de Chimie Mol\'eculaire et des Mat\'eriaux d'Orsay, Univ Paris Sud, Universit\'e Paris-Saclay, CNRS, UMR 8182, 91405 Orsay Cedex, France}

\date{\today}

\begin{abstract}
We  present a thorough   tight-binding  analysis  of the band structure of a wide variety of  lattices belonging to the class of honeycomb and Kagome  systems including several mixed forms combining both lattices. The band structure of these systems are made of a combination of dispersive and flat  bands. The dispersive bands possess Dirac cones (linear dispersion) at the six corners (K points) of the Brillouin zone although in peculiar cases  Dirac cones at the center of the zone $(\Gamma$ point) appear. The flat bands can be of different nature. Most of them are tangent to the dispersive bands at the center of the zone but some, for symmetry reasons, do not hybridize with other states.  The objective of our work is to provide an analysis of a wide class of so-called ligand-decorated honeycomb Kagome lattices that are observed in 2D metal-organic framework (MOF) where the ligand occupy honeycomb sites and the metallic  atoms the Kagome sites. We show that the $p_x$-$p_y$ graphene model  is relevant  in these systems  and there exists four types of flat bands: Kagome flat (singly degenerate) bands, two kinds of ligand-centered flat bands (A$_2$  like and E like, respectively doubly and singly degenerate) and metal-centered  (three fold degenerate) flat bands.

\end{abstract}

\maketitle

\section{Introduction}
\label{intro}

The discovery of the extraordinary electronic properties of graphene\cite{Novoselov2004} catalyzed the emergence of a new area of research the so-called two-dimensional (2D) materials. The number of reports on 2D systems both in the field of chemistry and physics exploded during the last ten years.
 The specificity of graphene that is at the origin of  its particular electronic transport properties  is related to the linear dispersion of some electronic bands in the vicinity of the Fermi level  forming Dirac cones at each corner of the Brillouin zone\cite{CastroNeto2009}. However graphene is a semi-metal {\sl i.e.} a zero-gap semiconductor which prevents using it in electronic devices. Since the seminal work on graphene many 2D inorganic materials that behave like semiconductors have been obtained such as the class of two-dimensional transition metal dichalcogenides\cite{Wang2012}. These materials are not new since their bulk properties were known for a long time\cite{Wilson1969} but the preparation of single layers led to the discovery of new electronic properties\cite{Mak2010} with potential applications in different fields of chemistry\cite{Xiaoyun2015} and physics\cite{Lee2012}. Recently, similarly to graphene, allotropes of silicon (sillicene), germanium (germanene),  tin (stanene) and phosphorus (phosphorene) have been obtained \cite{Balendhran2015}. Because of a larger spin-orbit coupling than in graphene these materials are predicted to be 2D topological insulators.\cite{Kou2015,Shirasawa2015}

A very interesting new class of materials is the Metal-Organic Frameworks (MOFs) that are obtained from the assembly of metal ions and organic ligands\cite{Batten2013}. A judicious choice of the organic
ligands and of  the coordination sphere of the metal ions can lead to the design of two-dimensional structures (2D MOF) with honeycomb Kagome arrangement \cite{Kambe2013,Kambe2014} where coordination bonds spread only in two dimensions and weak van der Waals interactions ensure the cohesion of the
solid in the third dimension\cite{Tsukamoto2017}. Depending on the electronic structures of the metal ions and of the organic ligands and on their possible interaction, materials with a large range of physical properties can be conceived and designed\cite{Kambe2014,Sun2015,Sun2016}. Despite several theoretical works essentially based on Density Functional Theory (DFT) calculations\cite{Zhao2013,Zhou2015,Wu2016,Yamada2016,Silveira2017} very little is known on the electronic structure of these promising materials. In particular the mechanism of formation of bands remains to be elucidated. Among other things it is crucial to analyze the nature of the dispersion relations:  Existence of ``slow'' or ``fast''  Dirac cones\cite{Wu2016}, existence of flat bands and interaction between bands, etc.

In this work, based on model tight-binding hamiltonians we derive analytically the energy dispersion of lattices of increasing complexity: starting from the well known graphene and kagome lattice we then explore mixed lattices (honeycomb-Kagome) as well as the so-called $p_x$-$p_y$ graphene model\cite{Wu2008} to end up   with a detailed analysis of the electronic structure of a generic MOF. 
\section{Geometrical description of lattices}
\label{geom-lattice}

\subsection{Triangular  lattice}
\label{triangular-lattice}

In this work we investigate various bi-dimensional lattices all belonging to the hexagonal (or triangular) Bravais lattice generated by the  translation vectors  ${\bm a}_1=a(-\frac{1}{2},\frac{\sqrt{3}}{2})$ and ${\bm a}_2=a(-\frac{1}{2},-\frac{\sqrt{3}}{2})$. For symmetry reasons, it is also convenient to introduce the vector ${\bm a}_3=-({\bm a}_1+{\bm a}_2)=a(1,0)$. The reciprocal lattice is hexagonal and its two reciprocal lattice vectors are   ${\bm b}_1=\frac{2\pi}{a}(-1,\frac{1}{\sqrt{3}})$ and ${\bm b}_2=\frac{2\pi}{a}(-1,-\frac{1}{\sqrt{3}})$. The first Brillouin zone is therefore an hexagon whose vertices are points $K=\frac{1}{3}({\bm b}_1+{\bm b}_2)$  and $K'=-K$  ($K$ and $K'$ are inequivalent in the case of graphene) and the ones obtained by  rotation of  $ \pm 2\pi/3$  . 

\begin{figure}[htbp]
\subfloat[ Bravais lattice]{\includegraphics[width=7cm]{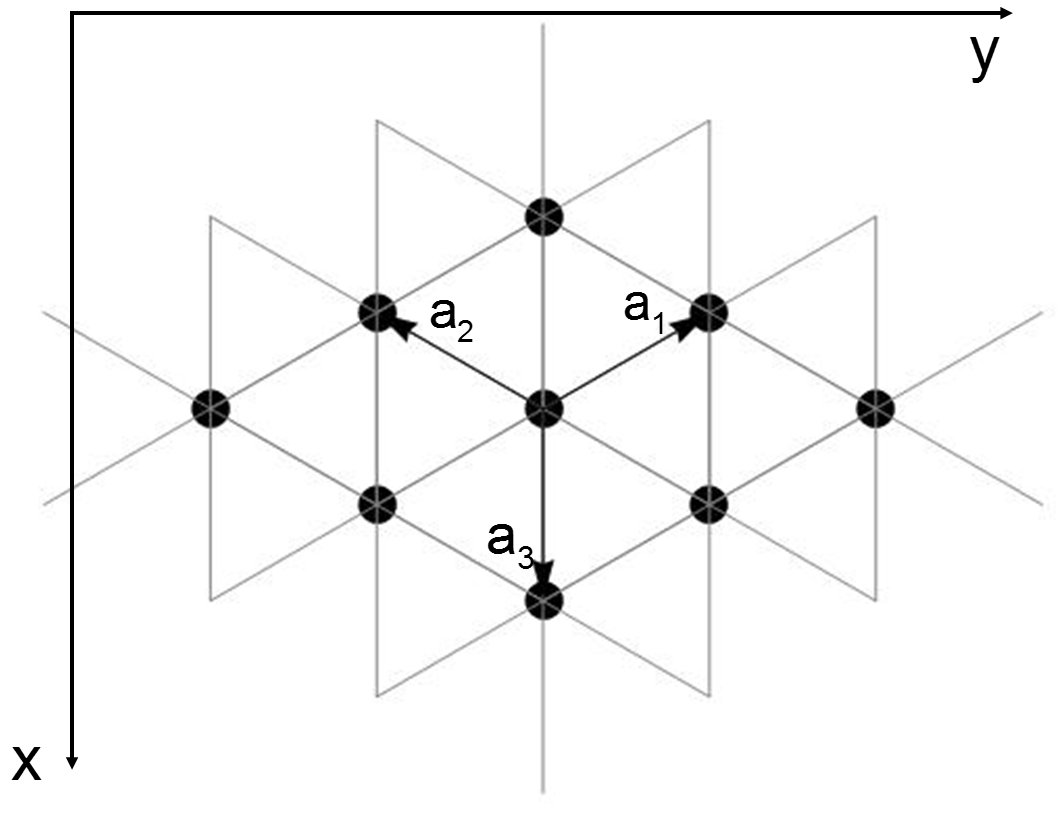}}
\subfloat[Reciprocal lattice]{\includegraphics[width=7cm]{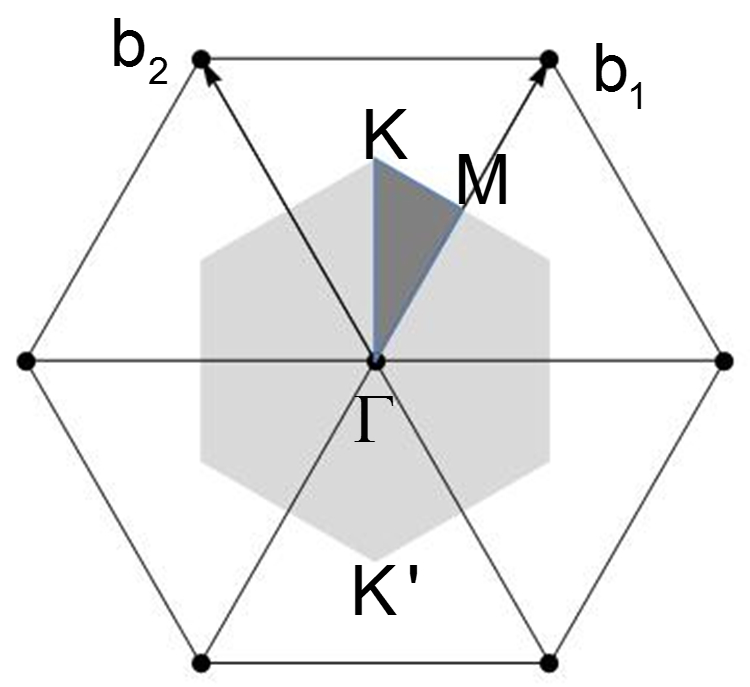}}
	\caption{\label{fig:triangular-lattice} a) Triangular Bravais lattice defined by its two translation vectors  ${\bm a}_1$ and ${\bm a}_2$ and  b)   its reciprocal lattice, defined by the vectors  $\bm b_1$ and $\bm b_2$. 
	The Brillouin zone is the light gray shaded hexagon and the irreducible Brillouin zone is  the dark gray shaded triangle with its three vertices: $\Gamma$, $K=\frac{1}{3}({\bm b}_1+{\bm b}_2)$ and $M=\frac{1}{2}{\bm b}_1$.  }
\end{figure}

\subsection{Honeycomb  lattice}
\label{honeycomb-lattice}

\begin{figure}[h!]
\onecolumngrid
\centering
	\includegraphics[width=7cm]{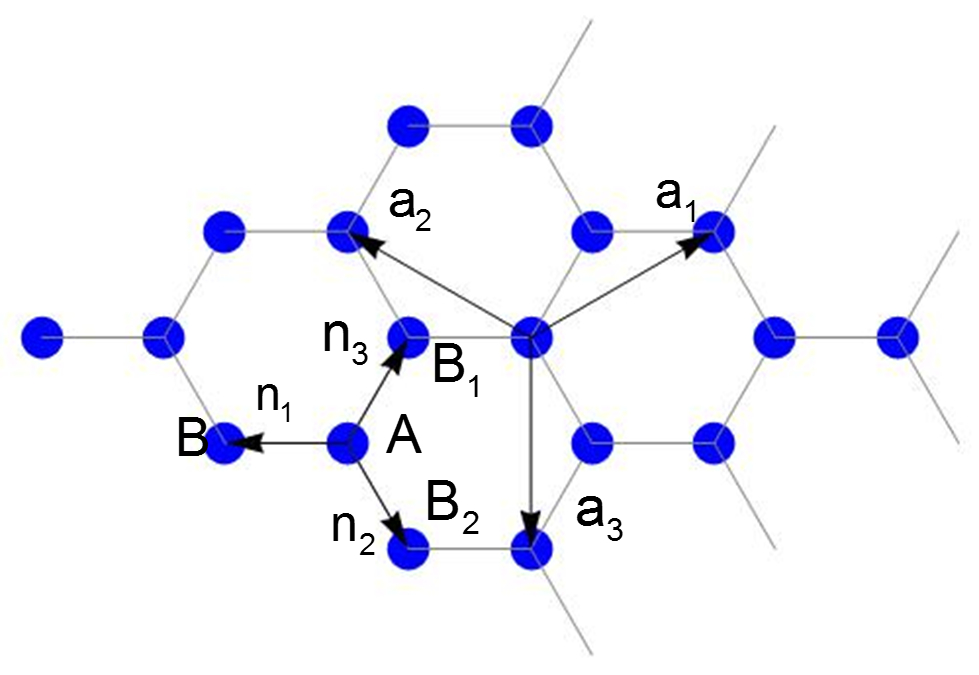}
	\caption{\label{fig:honeycomb-lattice} Honeycomb lattice where $\bm a_1$ and $\bm a_2$ are the two lattice vectors and $\pm {\bm n}_1$,  $\pm {\bm n}_2$ and  $\pm{ \bm n}_3$  are the three vectors connecting atoms from one sub-lattice (A or B) to the other (B or A). For example, atom A has three nearest neighbors: B, B$_1$=B+${\bm a}_1$ and B$_2$=B-${\bm a}_2$ . The elementary unit-cell contains 2 atoms (A and B).}
\end{figure}

The honeycomb lattice is not a  Bravais lattice itself but it is built from the hexagonal network decorated by two atoms (A and B) as illustrated in Fig.\ \ref{fig:honeycomb-lattice}. Each atom of the A sub-lattice  has three nearest neighbours from the sub-lattice B at distance $a/\sqrt{3}$  in directions ${\bm n}_1$,  ${\bm n}_2$ and  ${ \bm n}_3$ while on lattice B the connecting vectors are 
 $-{\bm n}_1$,  $-{\bm n}_2$ and  $-{ \bm n}_3$, where  ${\bm n}_i=\frac{1}{3}({\bm a}_{i+1}-{\bm a}_i)$, and reciprocally ${\bm a}_i=({\bm n}_{i+2}-{\bm n}_i)$. The most emblematic example of a honeycomb lattice is graphene which is a two-dimensional crystalline allotrope of carbon. Two-dimensional hexagonal boron nitride (h-BN) is another example but in that case the sub-lattices A and B are occupied by boron and nitrogen, respectively.
 
 The geometry of the honeycomb lattice being quite specific, we now define some quantities and demonstrate relations among them that will be useful in the analysis of the band structure.
 Let us first introduce the normalized vectors  $\bm{\hat{n}}_i=\frac{{\bm n}_i}{n_i}$ where $n_i=\Vert  {\bm n}_i \Vert$. The scalar product between two such vectors satisfies the relation $\bm{\hat{n}}_i.\bm{\hat{n}}_j=\frac{3}{2}\delta_{ij}-\frac{1}{2}$ and if one knows the scalar products of any vector $\bm v$ with the three linearly dependent vectors  $\bm{\hat{n}}_i$  ( $\sum_i \bm{\hat{n}}_i=0$)  it comes that:
 \begin{equation}
 {\bm v}=\frac{2}{3}\sum_i ({\bm v}_i.\bm{\hat{n}}_i)\bm{\hat{n}}_i \; .
 \end{equation}
It is also useful to define the three vectors:
\begin{equation}
\bm{u}(\bm{k}) = \sum_i \bm{\hat{n}}_i \,e^{i\bm{k}.\bm{n}_i} \quad ; \quad
\bm{u}^+(\bm{k})  = \sum_i \bm{\hat{n}}_i \,e^{i\bm{k}.\bm{n}_{i+1}} \quad ; \quad
\bm{u}^-(\bm{k})  = \sum_i \bm{\hat{n}}_i \,e^{i\bm{k}.\bm{n}_{i-1}}   \label{eq:def-u} \; ,
\end{equation}
where we have $\bm{u}+\bm{u}^++\bm{u}^- =0$ and $\bm{u}.\bm{u}^+=\bm{u}.\bm{u}^-=-1/2\bm{u}.\bm{u}$. Taking the scalar product of $\bm{u}$, $\bm{u}^+$ and $\bm{u}^-$ with   $\bm{\hat{n}}_i$ one gets:
 
  \begin{equation}
 {\bm u}.\bm{ \hat{n}}_i =\frac{3}{2} \,e^{i\bm{k}.\bm{n}_i} -\frac{1}{2} \gamma(\bm{k}) \quad ; \quad  {\bm u}^+.\bm{ \hat{n}}_i =\frac{3}{2} \,e^{i\bm{k}.\bm{n}_{i+1}} -\frac{1}{2} \gamma(\bm{k})
  \quad ; \quad  {\bm u}^-.\bm{ \hat{n}}_i =\frac{3}{2} \,e^{i\bm{k}.\bm{n}_{i-1}} -\frac{1}{2} \gamma(\bm{k})  \; ,
 \end{equation}
where $ \gamma( {\bm k})= e^{i{\bm k}.{\bm n}_1}+e^{i{\bm k}.{\bm n}_2}+e^{i{\bm k}.{\bm n}_3}$ is the sum of the nearest-neighbor phase factors that plays a central role in the band structure of the honeycomb and Kagome lattices. Let us derive expressions  in the vicinity of $\Gamma$ and $K$  which will be useful in the following:
\begin{align}
\gamma(\Gamma)&=3  \quad ; \quad   \gamma(\bm{q})\approx 3 -\frac{a^2}{4} (q_x^2+q_y^2)  \\
 \gamma(K)&= e^{iK.{\bm n}_1}+e^{iK.{\bm n}_2}+e^{i K.{\bm n}_3}=1+j^2+j=0 \quad ; \quad  \gamma(K+\bm{q}) \approx a \frac{\sqrt{3}}{2} (q_x-i q_y) \; ,
\end{align}
where $q=|\bm{q}| \ll 1/a$  and $j=e^{i\frac{2 \pi}{3}}$ (and $j^2=\bar{j})$ is the cubic root of unity. Note that in the vicinity of $K'=-K$ the expression is $\gamma(K'+\bm{q}) \approx -a\frac{\sqrt{3}}{2} (q_x+i q_y)$. For $ {\bm u}$ we have:
\begin{align}
\bm{u}(\Gamma) &=  \bm{u}^+(\Gamma)  = \bm{u}^-(\Gamma)  =0   \quad ; \quad   \bm{u}^+(\bm{q})  - \bm{u}^-(\bm{q}) \approx  i \sum_i  \bm{q}.(\bm{n}_{i+1}- \bm{n}_{i-1}) \bm{ \hat{n}}_i \\ 
i\bm{u}(K) &= \frac{3}{2}(1,-i)  \quad ; \quad \bm{u}^+(K)  = j^2  \bm{u}(K)  \quad ; \quad  \bm{u}^-(\Gamma)  =  j  \bm{u}(K)  \; , 
\end{align}
where one can note that  $\bm{q}.[\bm{u}^+(\bm{q})  - \bm{u}^-(\bm{q})] = 0$.

\subsection{Kagome lattice}
\label{kagome-lattice}

\begin{figure}[h!]
\onecolumngrid
	\centering
	\includegraphics[width=5cm]{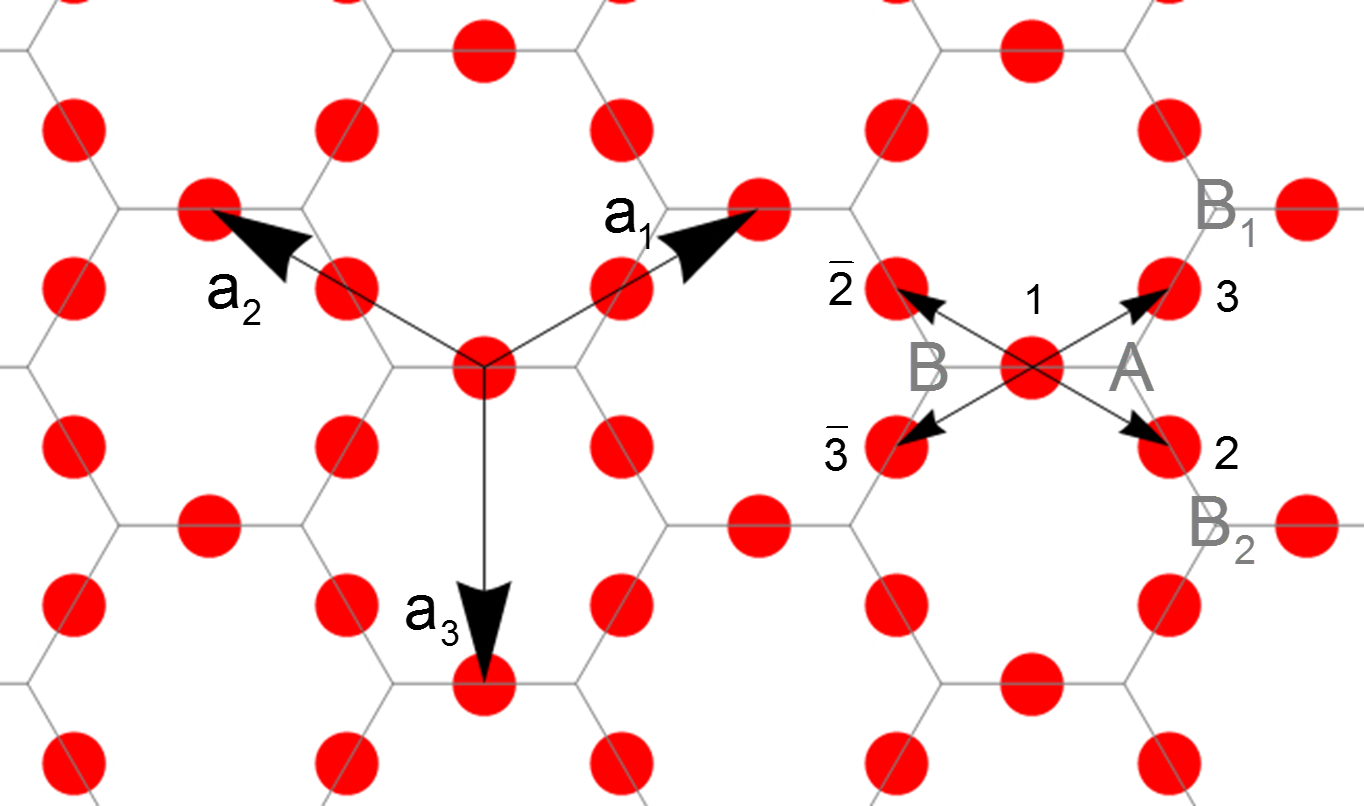}
	\caption{\label{fig:kagome-lattice} Kagome lattice where each atom has four nearest neigbors in directions $\pm{\bm a_i}/2$  The elementary unit-cell contains 3 atoms (1, 2 and 3).  The center of the triangle  formed by any three first neigbor atoms of the Kagome lattice form an honeycomb lattice labeled A and B. }
\end{figure}

The Kagome lattice \cite{Syozi1951}  is a structure that can be found in various minerals or molecular arrangements\cite{Johnston1990,Mekata2003}. It has attracted the attention of theoreticians due to its exotic electronic structure\cite{Bergman2008} (existence of flat bands)  and magnetic ordering\cite{Barros2014} (magnetic frustration). It can be obtained by decorating the honeycomb network by atomic sites in the middle of the hexagon edges as illustrated in Fig.\ \ref{fig:kagome-lattice}.  The Bravais lattice is still hexagonal with the same translation vectors but the unit cell is now made of three atoms (denoted 1, 2 and 3). Each atom has  four nearest neighbors at distance $a/2$ in directions $\pm{\bm a}_i/2$.

\subsection{Honeycomb-Kagome lattice}
\label{honeycomb-kagome-lattice}

\begin{figure}[h!]
\onecolumngrid
	\centering
	\includegraphics[width=6cm]{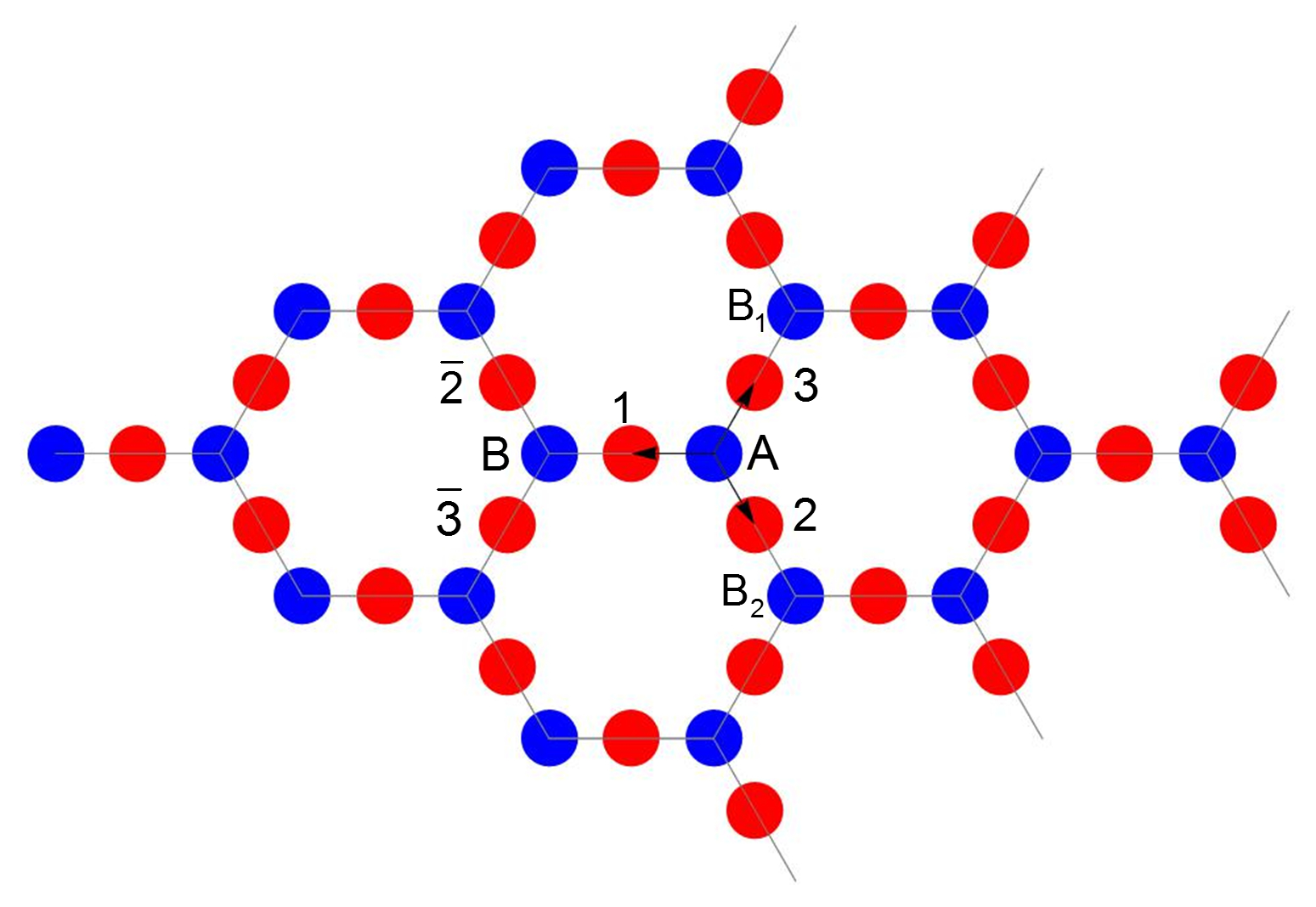}
	\caption{\label{fig:honeycomb-kagome} Honeycomb Kagome lattice where atoms from the honeycomb lattice are in blue and those on the Kagome lattice are in red.  The elementary unit cell contains five atoms (A, B, 1, 2, 3).}
\end{figure}

Let us now consider the lattice built from the superposition of the honeycomb and the Kagome lattice (hereafter denoted honeycomb-Kagome and first described in the seminal work by Syozi \cite{Syozi1951}). Each atom of the honeycomb lattice (in blue) has three nearest neighbors on the Kagome lattice while each  atom on the Kagome lattice (in red) has two nearest neigbors on the honeycomb lattice.

\subsection{$\alpha$-graphyne lattice}
 \label{graphyne-lattice}
 
 \begin{figure}[h!]
\onecolumngrid
	\centering
	\includegraphics[width=6cm]{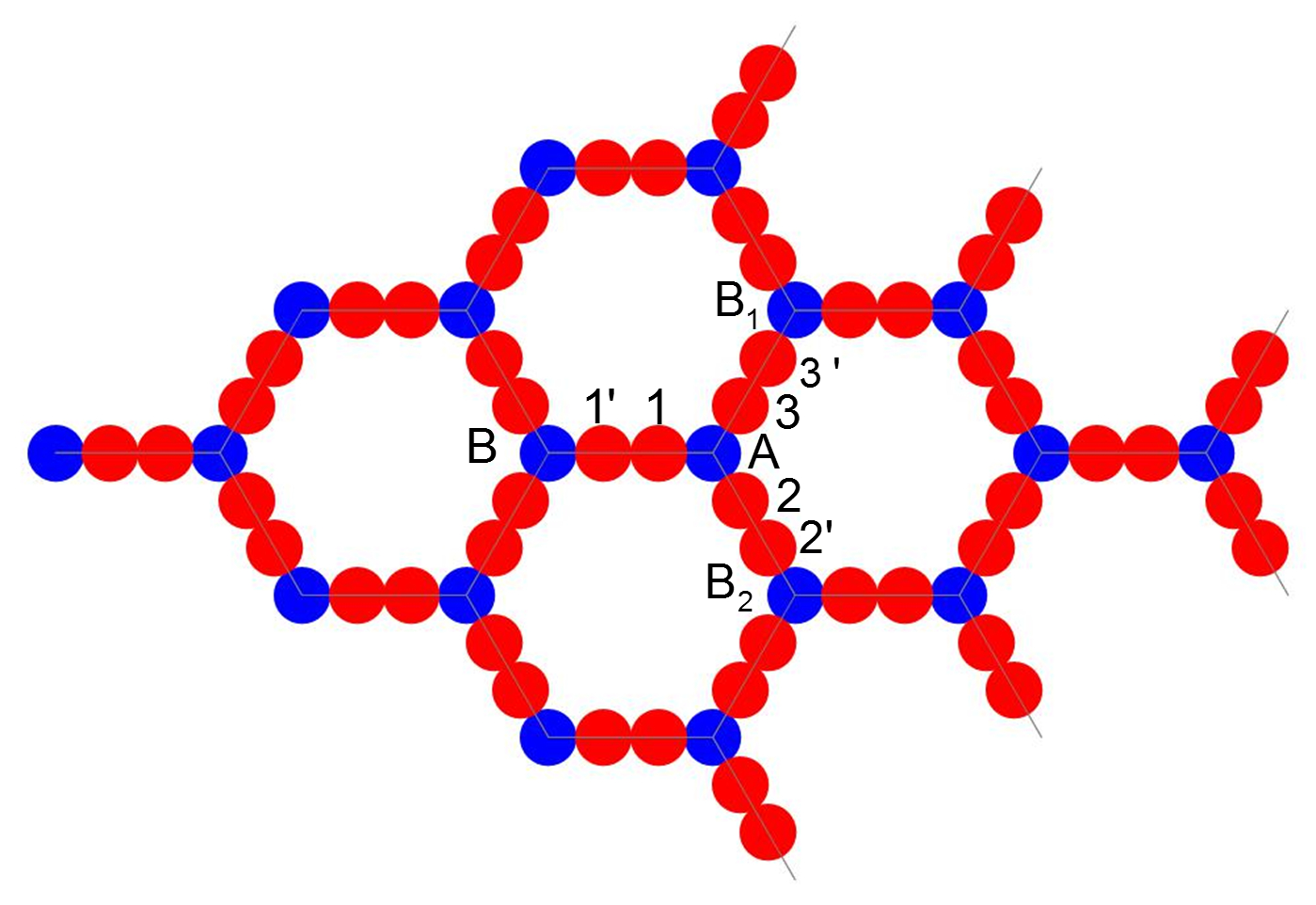}
	\caption{\label{fig:graphyne}$\alpha$-graphyne lattice. Its  elementary unit-cell contains eight atoms (A,B, 1, 1', 2, 2', 3, 3').}
\end{figure}

 The  2D $\alpha$-graphyne\cite{Baughman1987} is one of the many new allotropes of carbon that has been considered theoretically\cite{ZhihaiL2015} although it has not yet been synthesized.
$\alpha$-graphyne is an example of honeycomb-Kagome lattice but instead of decorating the edge of the hexagon by a single atom, two atoms are now located along each edge.  Atoms (in blue) occupying the honeycomb lattice have three nearest neighbors on the Kagome lattice, while each atom on the Kagome lattice (in red) has two nearest neighbors: one blue atom on the honeycomb lattice and one red atom on the Kagome lattice. The connecting vectors are of the form  $\pm{\bm n_i}/3$ (see Fig.\ \ref{fig:graphyne}).

\subsection{Ligand decorated honeycomb-Kagome lattice}
\label{decorated-honeycomb-kagome-lattice}

 \begin{figure}[ht]
\onecolumngrid
	\centering
	\includegraphics[width=12cm]{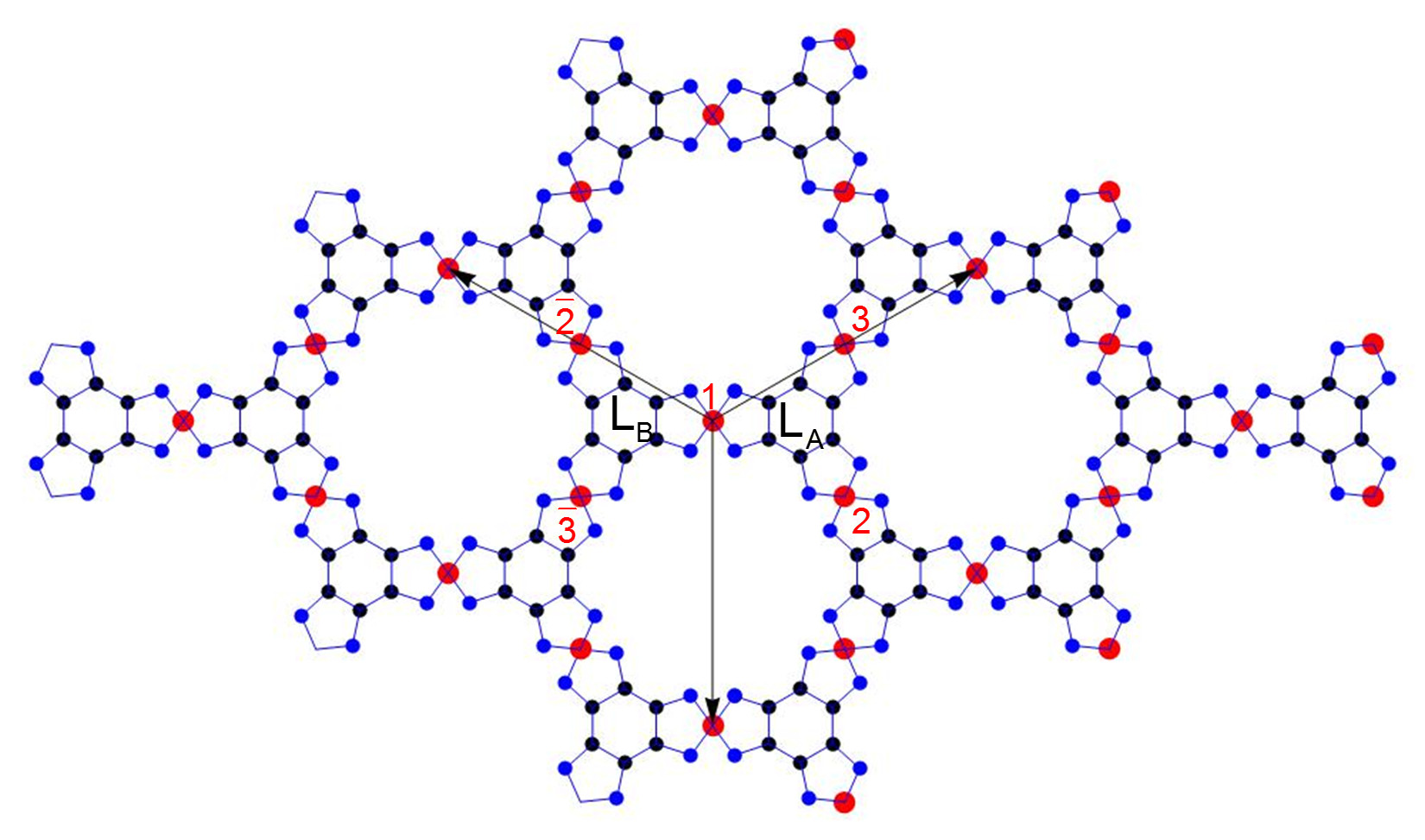}	
	\caption{\label{fig:decorated-honeycomb-kagome}Example of ligand-decorated honeycomb-Kagome lattice.  The elementary unit cell contains two identical ligands $L_A$ and $L_B$ occupying the honeycomb sites $A$ and $B$ and three atoms (1, 2 and 3) occupying Kagome sites. There are therefore $2n_L+3$ atoms in the unit cell.}
\end{figure}

Coordination chemistry offers a wide range of possibilities to design new types of 2D nanomaterials. In particular metal-organic frameworks  (MOF) are very attractive to implement  molecular arrangements whose electronic (and magnetic) properties  could be tailored. By combining organic ligands based on the combination of benzene-like rings (that possess a delocalized $\pi$-electrons system) with metallic ions, one can synthesize 2D coordination networks on an hexagonal lattice. Several of these proposed 2D MOF can be described as a structure in which the ligands occupy  a honeycomb lattice and are connected to  single ``bridge'' metallic ions occupying  Kagome sites. Hereafter these networks will be denoted ligand-decorated honeycomb-Kagome lattices. In addition we will suppose that the ligand is of $D_{3h}$ symmetry with rotation and mirror symmetries. 
If $n_L$ is the number of atoms in the ligand there are $2n_L+3$ atoms per unit cell.

\section{Electronic structure}
\label{electronic-structure}

In the following we investigate the electronic structure of the various lattices introduced in Sec.\ \ref{geom-lattice} within a simple tight-binding (TB) model including hopping integrals ($t$ or $t'$) between first neighbors only. Despite its simplicity,  we believe that the model can still capture the most important features and trends of more realistic models. We will derive whenever possible analytical formulae. Although the lattices considered are periodic, we will often use a real space expansion of the Schr\"{o}dinger equation in a manner similar to the seminal work by Thorpe, Weaire, Leman and Friedel to describe the electronic structure of sp$^3$ semiconductors\cite{Weaire1971,Leman1962}. This approach often proves to be more convenient than the traditional approach that consists in diagonalizing the Hamiltonian in $\bm{k}$ space {\sl i.e.} expressed in the basis  of TB Bloch states.

\subsection{$p_z$ graphene band structure}
\label{honeycomb-band}

Before considering the band structure of the honeycomb lattice, let us note that the energy dispersion $E(\bm{k})$ of the simple triangular lattice is simply given by:
\begin{equation}
E({\bm k})=2 t \,S({\bm k}) \quad \text{ with}  \quad  S({\bm k})= \cos {\bm k}.{\bm a}_1+\cos {\bm k}.{\bm a}_2+\cos {\bm k}.{\bm a}_3.
\end{equation}
The band structure of graphene (restricted to the subspace of $p_z$ orbitals) is obtained  by diagonalizing the tight-binding Hamiltonian $H({\bm k})$ in ${\bm k}$ space:
\begin{equation}
H({\bm k})=t\begin{pmatrix}
  0 &  \gamma( {\bm k}) \\
 \gamma^{\star}( {\bm k})  & 0
 \end{pmatrix} \; .
\end{equation}
The eigenvalues are then straightforwardly derived:
\begin{equation}
\label{eq:band-graphene}
E({\bm k})=\pm t|\gamma( {\bm k})| \; ,
\end{equation}
and the corresponding eigenvectors are given by:
\begin{equation}
\label{eq:vec-graphene}
\begin{pmatrix}
  c_A( {\bm k}) \\
  c_B( {\bm k})
 \end{pmatrix}
 =
 \begin{pmatrix}
  e^{i  \theta_{\bm k}/2} \\
    \pm   e^{-i  \theta_{\bm k}/2}
 \end{pmatrix} \; ,
\end{equation}
where $\theta_{\bm k}$ is the argument of $\gamma( {\bm k})$ {\sl i.e.}  $\gamma( {\bm k})=|\gamma( {\bm k})|  e^{i  \theta_{\bm k}}$. \\

\begin{figure}[htbp]
\subfloat[ $\Delta=0$]{\includegraphics[width=7cm]{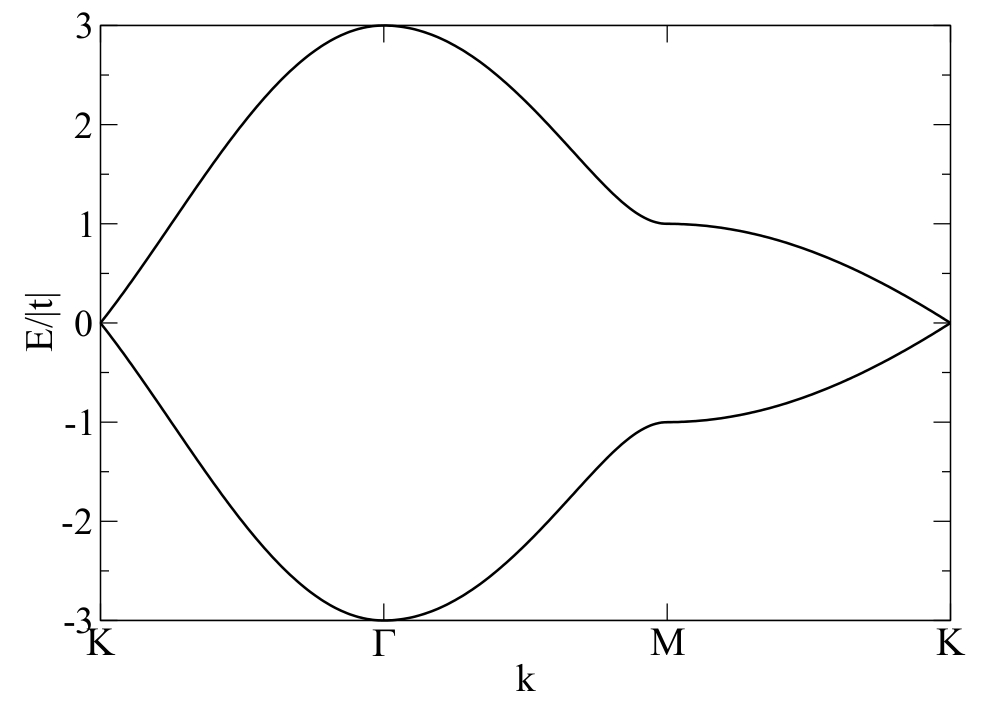}}
\subfloat[$\Delta\ne0$]{\includegraphics[width=7cm]{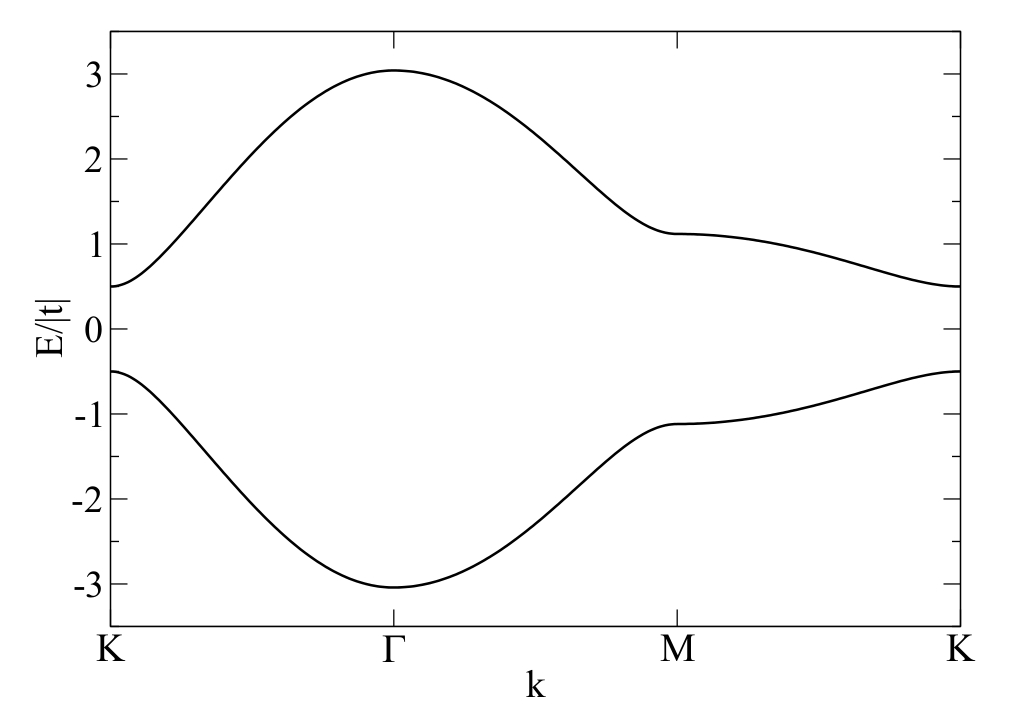}}
	\caption{\label{fig:graphene-band} Graphene band structure a) without  and b) with onsite  $\pm \Delta$}
\end{figure}

The specificity of the  graphene band structure is its linear dispersion (which makes the Bloch electrons massless) in the vicinity of the ${\bm k}=K(K')$ points of the Brillouin zone where $\gamma(K)=0$ . 
Indeed if we decompose the vector ${\bm k}=K+{\bm q}$ where $q<<1/a$  then, to  first order, we have seen that $\gamma(K+\bm{q}) \approx a \frac{\sqrt{3}}{2} (q_x-i q_y)$, and $E(K+{\bm q})\approx \pm t \frac{\sqrt{3}}{2}a \sqrt{q_x^2+q_y^2}= \pm \hbar v_F q$ while in the vicinity of the $\Gamma$ point one has $\gamma({\bm q})\approx 3-\frac{a^2}{4} (q_x^2+q_y^2)$.
In addition, in neutral graphene the Fermi level is located right in the middle of the spectrum where the upper band $\pi^{\star}$ touches the lower band $\pi$ at the six vertices of hexagonal Brillouin zone ($K$ and $K'$ points). The dispersion relation along the K$\Gamma$MK path is shown in Fig.\ \ref{fig:graphene-band}a where the linear behavior is clearly visible around the K point where the Hamiltonian can be approximated by:
\begin{equation}
H(K+{\bm q})=|t| a \frac{\sqrt{3}}{2}\begin{pmatrix}
  0 &  (q_x-i q_y)\\
 (q_x+i q_y)  & 0
 \end{pmatrix}=
\hbar v_F \bm{q}.\bm{\sigma} \; ,
\end{equation}
where the components $\sigma_{x,y}$ of $\bm{\sigma}$ are the Pauli matrices. Close to K', since $\gamma(K'-\bm{q}) =\gamma(K+\bm{q})^{\star} $, we have $H(K'-{\bm q})=\hbar v_F (q_x,-q_y).\bm{\sigma}$.
 Let us now derive the dispersion relation by expressing the Schr\"{o}dinger equation in terms of the TB basis expansion coefficient $c_i$, where $i$ denotes the sites of the graphene lattice $\ket{\Psi}=\sum_i c_i \ket{i}$.  Each A atom has the same surrounding environment with three nearest neighbors B, B$_1$ and B$_2$ and therefore we have:
  \begin{equation}
  \label{eq:Shro-graphene-real}
  E({\bm k}) c_A=t(c_B+c_{B_1}+c_{B_2}) \; .
  \end{equation}
Making use of Bloch theorem, the following relations apply: $c_{B_2}(\bm{k})=e^{-i{\bm k}.{\bm a}_2}c_{B}(\bm{k})$ and  $c_{B_1}(\bm{k})=e^{i{\bm k}.{\bm a}_1}c_{B}(\bm{k})$ so that:
 \begin{equation}
  E({\bm k}) c_A(\bm{k})=t e^{-i{\bm k}.{\bm n}_1} (e^{i{\bm k}.{\bm n}_1}+e^{i{\bm k}.{\bm n}_2}+e^{i{\bm k}.{\bm n}_3})c_{B}(\bm{k})=t e^{-i{\bm k}.{\bm n}_1} \gamma( {\bm k})c_{B}(\bm{k})  \; ,
  \end{equation}
 and similarly for site B:
    \begin{equation}
  E({\bm k}) c_{B}(\bm{k})=t e^{i{\bm k}.{\bm n}_1} \gamma^{\star}( {\bm k})c_{A}(\bm{k})  \; .
  \end{equation}
The combination of both equations leads to:
      \begin{equation}
    E^2({\bm k}) c_{i}(\bm{k})=t^2|\gamma( {\bm k})|^2 c_{i}(\bm{k}) \quad ; \quad i=A,B \; ,
    \end{equation}
and we recover the energy dispersion  of Eq. (\ref{eq:band-graphene}).\\

When considering the case where  sites A and B are occupied by atoms of different nature, a straightforward extension of the TB model consists in adding a varying on-site energy depending on the occupation of this site by an A or a B atom. Let us define on-site levels $\epsilon_i=\sigma_i \Delta$ where $\sigma_i=1$ if site $i$  belongs to sublattice A and $\sigma_i=-1$ if site $i$  belongs to sub-lattice B. The Hamiltonian now reads:
\begin{equation}
H({\bm k})=\begin{pmatrix}
  \Delta &  t \gamma( {\bm k}) \\
t  \gamma^{\star}( {\bm k})  & -\Delta
 \end{pmatrix}  \; ,
\end{equation}
and the corresponding dispersion relation is:
\begin{equation}
\label{eq:band-BN}
E({\bm k})=\pm \sqrt{\Delta^2+t^2|\gamma( {\bm k})|^2}  \quad \text{ with}  \quad  |\gamma( {\bm k})|^2=3+2S( {\bm k})  \; .
\end{equation}
A direct gap $2 \Delta$ is now opened at point K and the linear dispersion is lost.  This typical situation occurs in systems such as hexagonal boron nitride.

\subsection{Kagome band structure}
\label{band-kagome}

The band structure of the Kagome lattice is slightly more complicated to derive\cite{ Liu2014}. Let us start with the Hamiltonian $H({\bm k})$ in $\bm k$ space:
\begin{equation}
H({\bm k})=t\begin{pmatrix}
  0               &  2 \cos k_2 &  2 \cos k_1\\
2 \cos k_2   & 0  &  2 \cos k_3  \\
2 \cos k_1 & 2 \cos k_3 &   0
 \end{pmatrix}
 =
\begin{pmatrix}
  0    & t_2 &  t_1\\
t_2   & 0   & t_3  \\
t_1 & t_3  &   0  
 \end{pmatrix} \; ,
\end{equation}
where $ k_{\nu}=\frac{1}{2}{\bm k}.{\bm a}_{\nu}$ and $ t_{\nu}=2t \cos (\frac{1}{2}{\bm k}.{\bm a}_{\nu})$.
The characteristic polynomial of the Hamiltonian matrix is given by $P(\lambda)=\det(\lambda I -H)=\lambda^3-(t_1^2+t_2^2+t_3^2)\lambda -2t_1t_2t_3$.
Using the relations $\prod_{\nu=1}^{3} t_{\nu}=2t^3(1+S)$,  $\sum_{\nu=1}^{3} t_{\nu}^2 = 2t^2(3+S)$ and $|\gamma ({\bm k})|^2=3+2S$,  the characteristic polynomial can be factorized $P(\lambda)=(\lambda+2t)((\lambda-t)^2-t^2|\gamma ({\bm k})|^2)$ and the energy dispersion reads:
\begin{equation}
E_{1,2}=t(1 \pm |\gamma ({\bm k}|)) \quad ; \quad E_{3}=-2t  \; .
\end{equation}
\begin{figure}[htbp]
\subfloat[ $t<0$]{\includegraphics[width=7cm]{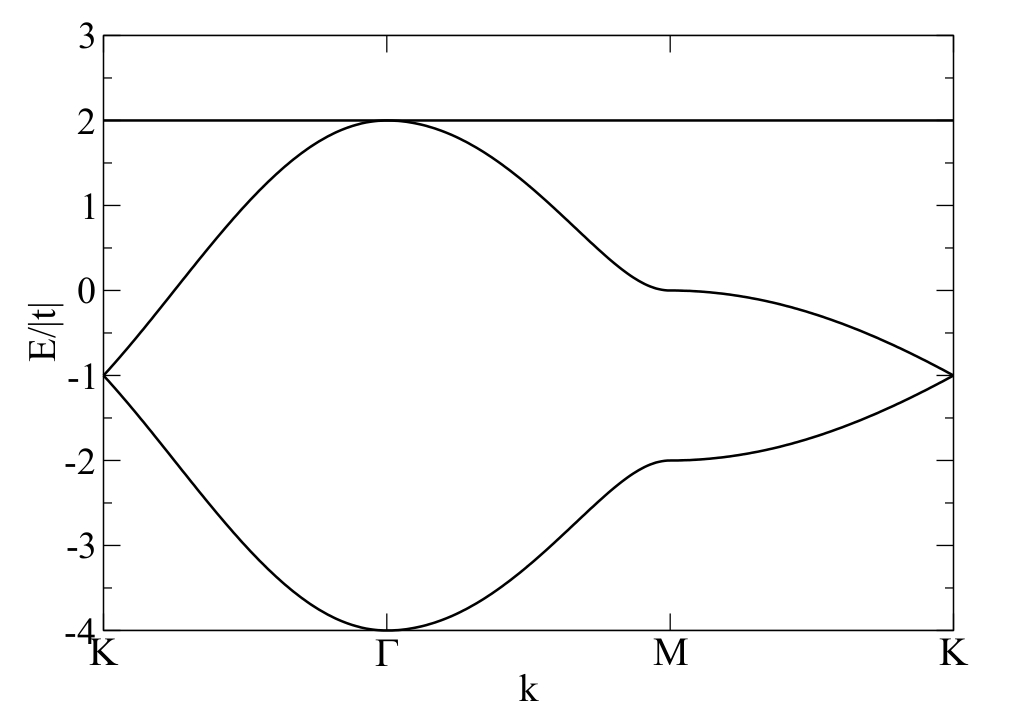}}
\subfloat[$t>0$]{\includegraphics[width=7cm]{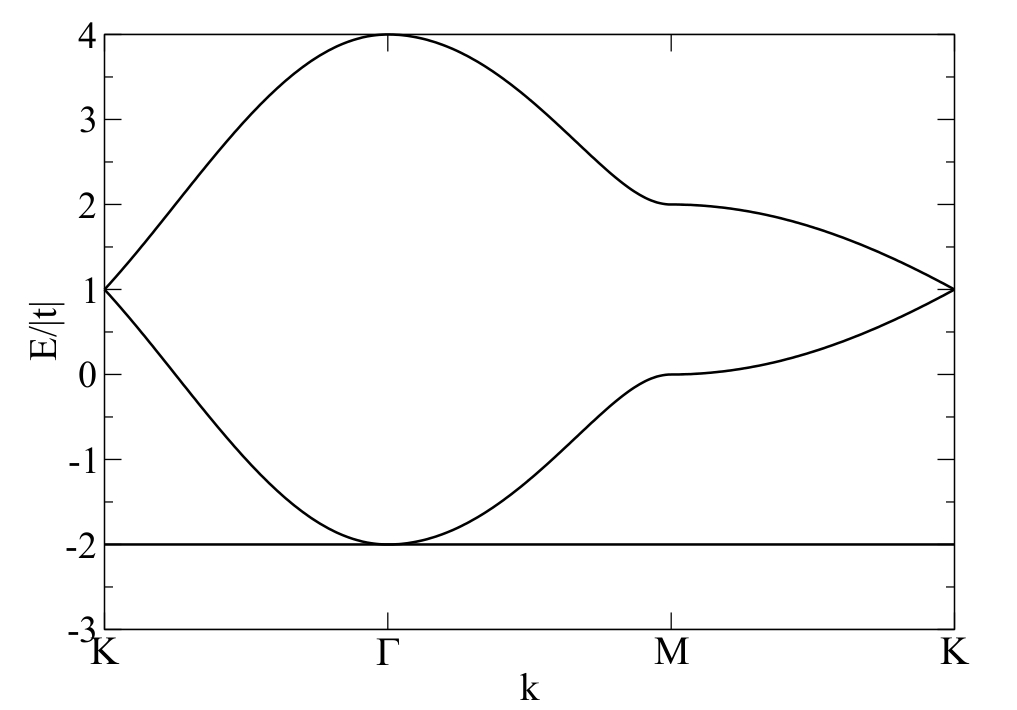}}
	\caption{\label{fig:kagome-band} Kagome band structure with a) negative and b) positive hopping  }
\end{figure}
The Kagome band structure is therefore composed of two branches ($E_{1,2}({\bm k})$) similar to those of graphene with an additional non dispersive branch $E_3$ touching the top or the bottom of the dispersive branch (depending on the sign of the hopping integral $t$) at the center of the Brillouin zone as illustrated in Fig.\ \ref{fig:kagome-band}. 
This derivation of the dispersion relation based on a direct diagonalization of  $H({\bm k})$  is rather cumbersome and we  now propose another approach which, in addition, has the advantage of allowing a straightforward derivation of  the eigenvectors. First one can start from the expression of $H({\bm k})$ whose components can be expressed as $h_{i,j}(\bm{k})=2 t ( \cos[\frac{\bm k}{2}.({\bm n}_i-{\bm n}_j)] -\delta_{ij})$.  The eigenstate equation now reads:
\begin{equation}
E\, c_i(\bm{k})=\sum_j  h_{ij} c_j(\bm{k})=\sum_j \big [ e^{i \frac{\bm k}{2}.({\bm n}_i-{\bm n}_j)}+ e^{-i \frac{\bm k}{2}.({\bm n}_i-{\bm n}_j)} -2\delta_{ij} \big] t\, c_j (\bm{k}) ,
\end{equation}
or else
\begin{equation}
\label{eq:kagome-band2}
(E+2t) c_i(\bm{k})=  e^{-i \frac{\bm k}{2}.{\bm n}_i}t \xi_{\bm k}+  e^{i \frac{\bm k}{2}.{\bm n}_i}t \xi_{-\bm k} \quad \text{where} \quad  \xi_{\bm k}=\sum_i   e^{i \frac{\bm k}{2}.{\bm n}_i} c_i(\bm{k}).
\end{equation}
Multiplying by $e^{\pm i \frac{\bm k}{2}.{\bm n}_i}$ and summing over $i$ gives:
\begin{equation}
t \begin{pmatrix}
 1                                     &   \gamma({\bm k}) \\
 \bar{\gamma}({\bm k})  &1 
 \end{pmatrix}
\begin{pmatrix}
\xi_{\bm k} \\
\xi_{-\bm k}
 \end{pmatrix} =E
 \begin{pmatrix}
\xi_{\bm k} \\
\xi_{-\bm k}
 \end{pmatrix}  \; .
\end{equation}
Hence $E=t(1 \pm |\gamma({\bm k}))|$ except if for all $ {\bm k},  \, \, \xi_{\bm k}= \xi_{-\bm k}=0 $, in which case we have from Eq.\ (\ref{eq:kagome-band2}) $E=-2t$. One can check that $c_i(\bm{k})= \sin[\frac{\bm k}{2}.({\bm n}_{i+1}-{\bm n}_{i+2})]=\sin k_{i+2} $ (\textit{i.e.} $(c_1,c_2,c_3)=(\sin k_3,\sin k_1, \sin k_2)$ ) is the eigenfunction of the localized state:
\begin{align}
\sum_j h_{ij} c_j(\bm{k})&=2\sum_j \cos[\frac{\bm k}{2}.({\bm n}_{i}-{\bm n}_{j})]   \times  \sin[\frac{\bm k}{2}.({\bm n}_{j+1}-{\bm n}_{j+2})] \nonumber \\                         
                              &=2\sum_j \sin[\frac{\bm k}{2}.({\bm n}_{j+1}-{\bm n}_{j+2}+ {\bm n}_{i}-{\bm n}_{j}  )]   + \sin[\frac{\bm k}{2}.({\bm n}_{j+1}-{\bm n}_{j+2}+ {\bm n}_{j}-{\bm n}_{i}  )] =0  \;
\end{align}
since ${\bm n}_{j}+{\bm n}_{j+1}+ {\bm n}_{j+2}=0$. The two other eigenstates are given by $c_i(\bm{k})=\cos \frac{1}{2}({\bm k}.{\bm n}_i-\theta_k)$ and  $c_i(\bm{k})=\sin \frac{1}{2}({\bm k}.{\bm n}_i-\theta_k)$ for $E=t(1+|\gamma({\bm k})|)$ and $E=t(1-|\gamma({\bm k})|)$, respectively, $\theta_k$ being the argument of $|\gamma({\bm k})|$ (see Eq. \ref{eq:vec-graphene}). In $\Gamma$ the dispersive band $E=t(1-|\gamma({\bm k})|)$ and the flat band are touching and the eigenvalue $-2t$ is doubly degenerate. The corresponding  eigenstates belong to the bidiemensional space  orthogonal to the eigenstate $(1,1,1)$ of eigenvalue $E=t(1+|\gamma({\bm k})|)$ .\\

\begin{figure}[h]
\onecolumngrid
	\centering
	\includegraphics[width=5cm]{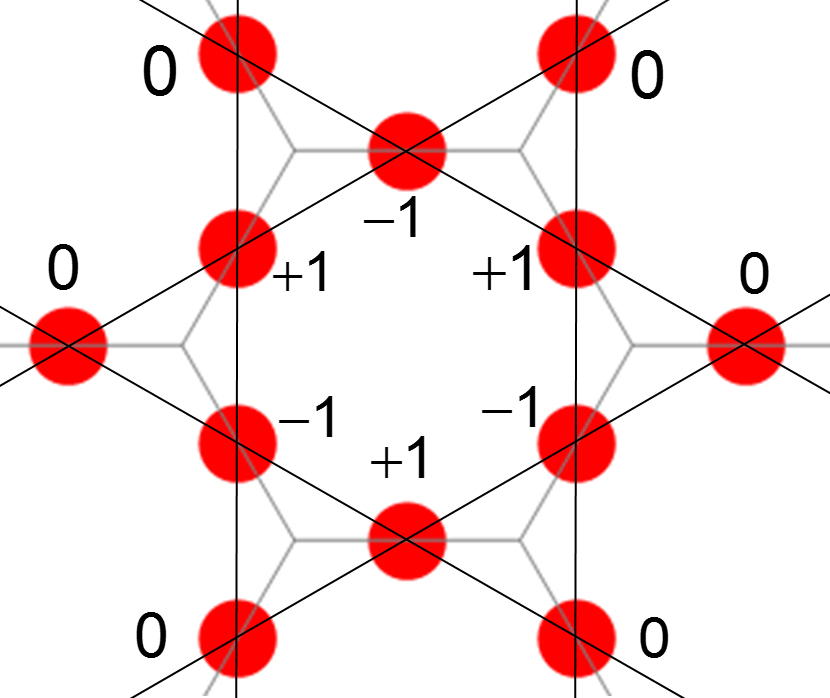}
	\caption{\label{fig:kagome-flat} Schematic representation of the Kagome flat band localized states ($E=-2t$)}
\end{figure}

The existence of dispersionless states is specific to the Kagome lattice but can appear in other lattices as well\cite{Liu2014}. It is related to the existence of destructive interferences which however are not obviously seen from the expression of the associated  Bloch states $(\sin k_3,\sin k_1, \sin k_2)$. In order to obtain a spatial description of  the  localized states associated to the flat band it is more convenient to express Schr\"{o}dinger equation in real space. Starting from site 1 (see Fig.\ \ref{fig:kagome-lattice}) that has four nearest neigbors, the following relation holds:
\begin{equation}
\label{eq:kagome-Shro-real}
E c_1=t(c_2+c_3+c_{\bar{2}}+c_{\bar{3}})=t(C_A+C_B-2c_1),
\end{equation}
where by definition we have set $C_A=c_1+c_2+c_3$ and $C_B=c_1+c_{\bar{2}}+c_{\bar{3}}$. Similarly we can derive equations for sites 2 and 3, $E c_2=t(C_A+C_{B_2}-2c_2)$ and $E c_3=t(C_A+C_{B_1}-2c_3)$. Summing up these three equations one obtains:
\begin{equation}
(E-t) C_A=t(C_B+C_{B_1}+C_{B_2})  \; ,
\end{equation}
which, apart from an additional onsite shift $t$, is exactly similar to Eq.\ (\ref{eq:Shro-graphene-real}) for graphene and therefore one  recovers the energy dispersion of the dispersive bands $E_{1,2}=t(1 \pm |\gamma ({\bm k}|))$ which holds only if the coefficients $C_i$ are different from zero. In case where all the coefficients $C_i$ vanish, and using Eq.\ (\ref{eq:kagome-Shro-real}) the flat band $E_{3}=-2t$ is also recovered. The associated eigenstates are such that the sum of their coefficients on any triangle of first neigbor atoms vanishes. It is thus possible to build a set of linearly independent eigenfunctions localized on hexagonal rings with alternate $\pm 1$ expansion coefficients such as the one schematically represented in Fig.\ \ref{fig:kagome-flat}. These localized states are exact eigenstates due to destructive interferences\cite{Bergman2008}. The eigenfunctions of two adjacent hexagons are not orthogonal due to their overlap but  it can be shown that the set of functions formed is linearly independent\cite{Bergman2008}.

\begin{figure}[htbp]
\subfloat[ $\Delta= t$]{ \includegraphics[width=7cm]{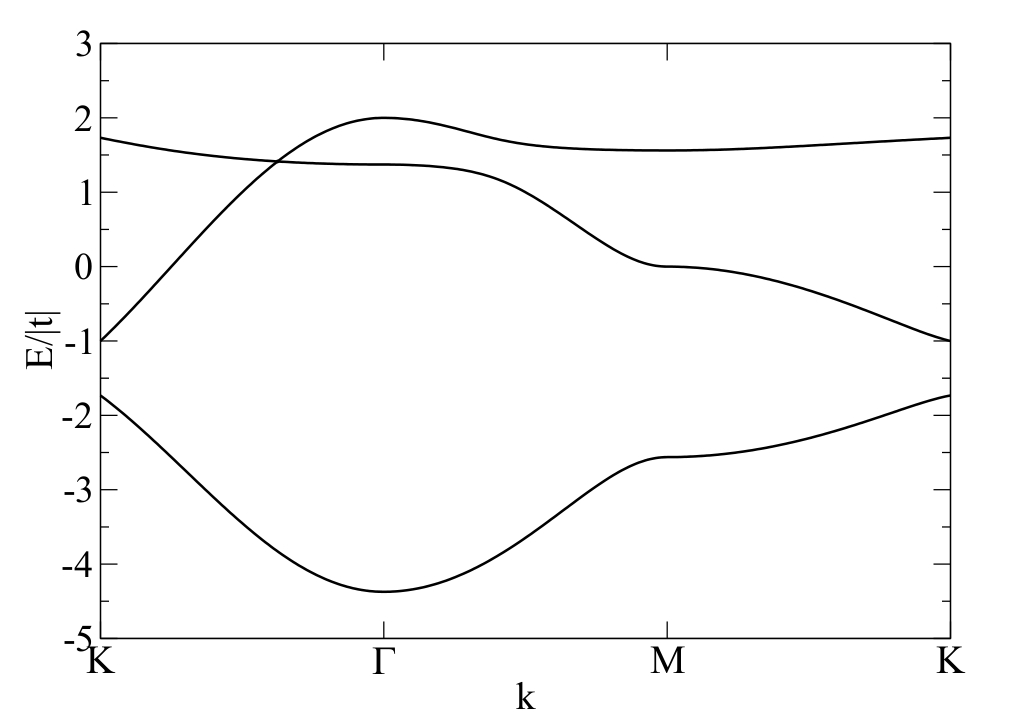}}
\subfloat[$\Delta= -t$]{\includegraphics[width=7cm]{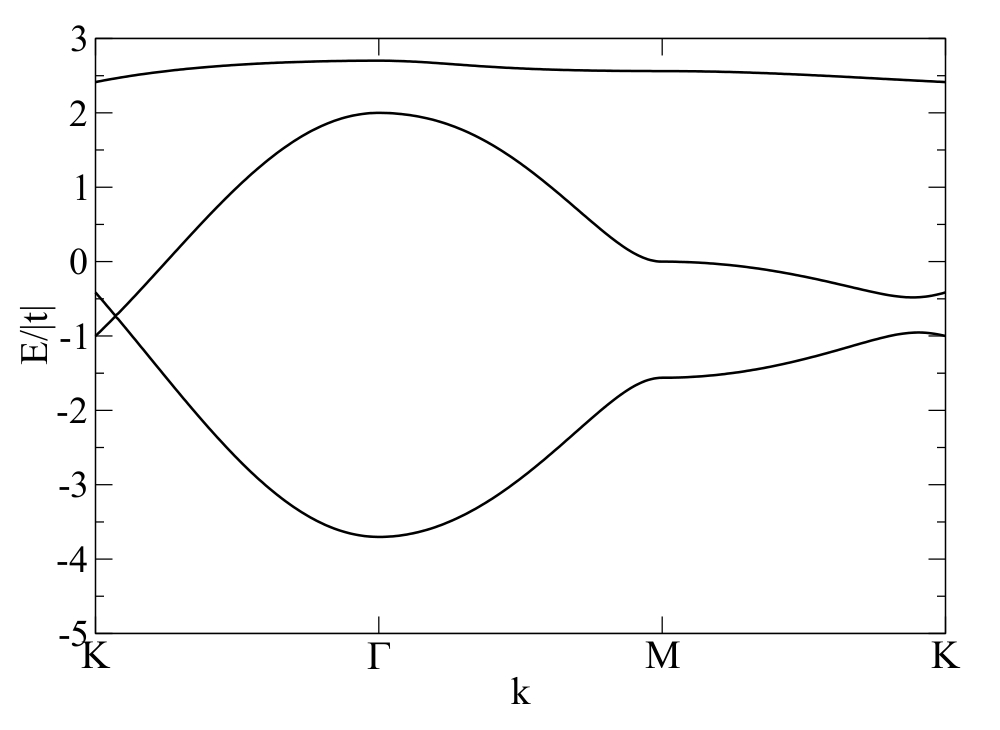}}
	\caption{\label{fig:kagome-band-onsite}Kagome band structure for a negative hopping $t$  and a)  negative  or b) positive on-site $\Delta=\pm t$ on atom 1}
\end{figure}

It is also important to note that the existence of these perfectly non-dispersive states is not robust: the inclusion of next nearest neighbor interactions induces a dispersion of these states\cite{Takeda2004} (with second neighbour interaction the electron can ``escape'' from the hexagonal ring), as well as  if one of the three atom of the Kagome unit cell is different from the other (which happens if one of the atom is of different nature or if its electronic or magnetic states differs from the other). For illustration we have plotted in Fig.\ \ref{fig:kagome-band-onsite} the band structure of a Kagome lattice in case where a positive (or negative) on-site $\Delta=\pm t$ ($t<0$) is added to the atom 1 of the unit cell.

\subsection{$p_x-p_y$ graphene band structure}
\label{px-py-honeycomb-band}

\begin{figure}[h!]
\onecolumngrid
	\centering
	\includegraphics[width=9cm]{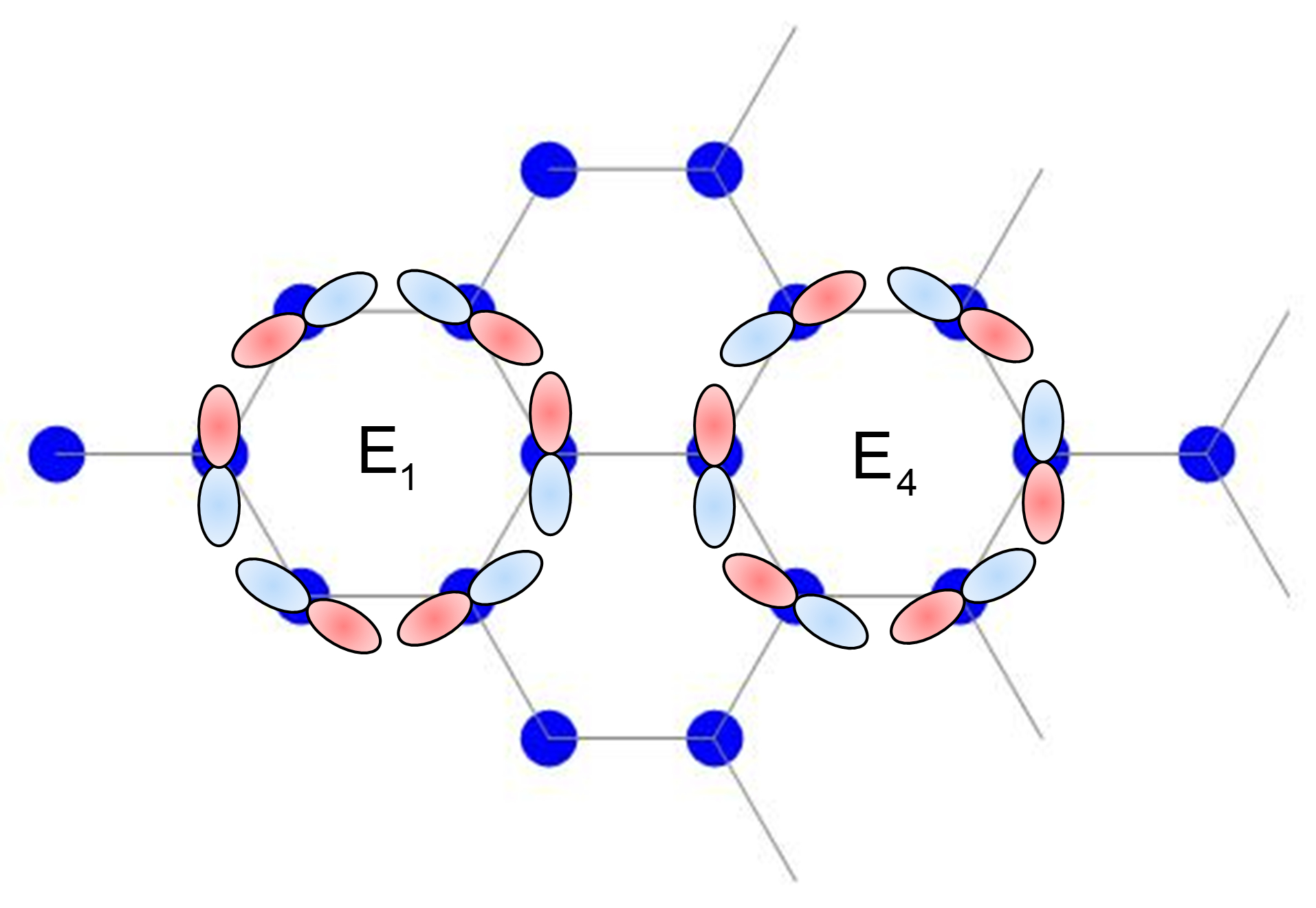}
	\caption{\label{fig:graphene-px-py-localized-states} Schematic representation of the localized eigenstates (E$_1$) and (E$_4$) corresponding to the flat bands $E=-3/2t$ and $3/2t$ respectively. Blue and pink lobes  represents the positive and negative parts of the $p$ orbitals respectively.}
\end{figure}

In  real graphene the orbitals can be split into two sets that do not hybridize: $sp^2$ (obtained as a linear combination of $s,p_x$ and $p_y$) and $p_z$. The $sp^2$ orbitals hybridize to form the bonding $\sigma$ and anti-bonding $\sigma^{\star}$ bands separated by a gap. The $p_z$ orbitals hybridize to form the $\pi$ and $\pi^{\star}$ bands described in the previous paragraph.  In the context of cold-atom physics,  Wu {\sl et al.} \cite{Wu2007,Wu2008} proposed a $p_x, p_y$ counterpart of graphene. We will see that this model is in fact relevant in the case of the ligand decorated honeycomb-Kagome lattice (Fig. \ref{fig:decorated-honeycomb-kagome}). In this model, only $p_x$ and $p_y$ are taken into account and $pp\pi$ Slater-Koster hopping integrals are neglected. For the sake of generality we will  consider two orbitals $p_{\lambda}$  and  $p_{\mu}$ pointing in the two orthogonal directions $\bm{\hat{\lambda}}$ and $\bm{\hat{\mu}}$ (with $\bm{\hat{\lambda}}.\bm{\hat{\mu}}=0$). The  hopping integral $t_{\lambda, \mu}^{i j}=\langle p_{i,\lambda}|H|p_{j,\mu} \rangle$ between two first neigbor sites $i$ and $j$  is simply given by $t_{\lambda, \mu}^{i j}=t(\bm{\hat{\lambda}}.\bm{\hat{n}}_{ij})(\bm{\hat{\mu}}.\bm{\hat{n}}_{ij})$ where $\bm{\hat{n}}_{ij}$ is the unit vector of the  $ij$ bond (\textit{i.e.} $\bm{\hat{n}}_{ij}= \bm{\hat{n}}_{j} \text{ if } i \in A, - \bm{\hat{n}}_{j} \text{ if } i \in B$) and  $t=pp\sigma$. 
\begin{equation}
E\,c_{i \lambda}  =  \sum_{j \mu} t_{\lambda, \mu}^{i j} c_{j \mu}  = t   \sum_{j}   (\bm{\hat{n}}_{ij}.\bm{\hat{\lambda}})  (\bm{\hat{n}}_{ij} .\sum_{\mu} c_{j \mu}\hat{\bm{\mu}}) \; .
\end{equation} 
Setting $\bm{C}_i=\sum_{\hat{\lambda}} c_{i \lambda} \bm{\hat{\lambda}}$ we then have the vectorial equation:
\begin{equation}
\label{eq:pxpx-honeycomb}
E\bm{C}_{i}  =  t \sum_{j } (\bm{\hat{n}}_{ij} .\bm{C}_{j} ) \bm{\hat{n}}_{ij}  \; .
\end{equation} 
Taking the scalar product and using the relation $\bm{\hat{n}}_{ij}.\bm{\hat{n}}_{ik}=3/2\, \delta_{kj}-1/2$ one gets:
\begin{equation}
E(\bm{C}_{i}.\bm{\hat{n}}_{ik})  =  \frac{3\,t}{2} (\bm{\hat{n}}_{ik}.\bm{C}_{k})
-\frac{t}{2}\sum_{j } (\bm{\hat{n}}_{ij}.\bm{C}_{j} ) \; .
\end{equation}
Setting $b_i=\sum_{j } (\bm{\hat{n}}_{ij} .\bm{C}_{j} ) $ we can derive two equations:
\begin{subequations}
\begin{align}
E\,(\bm{C}_{i}.\bm{\hat{n}}_{ik}) & -  \frac{3t}{2} ( \bm{C}_{k}. \bm{\hat{n}}_{ik} )= -\frac{t}{2} b_i  \label{eq1}  \\
  -  \frac{3t}{2} (\bm{C}_{i}. \bm{\hat{n}}_{ik}) & +  \,E(\bm{C}_{k}.\bm{\hat{n}}_{ik})= \frac{t}{2} b_k  \; . \label{eq2}
\end{align}
\end{subequations}
Summing in Eq.(\ref{eq1}) over the neighbors $i$ of site $k$ one gets:
\begin{equation}
E\,b_k  = \frac{t}{2} \sum_i b_i  \; .
\end{equation}
If the $b_i$ are not all zero the solutions are the ones of graphene:
\begin{equation}
E_{2,3}=\pm \frac{|t|}{2}|\gamma({\bm k})|  \; ,
\end{equation}
while if, for all $i,  \; b_i=0$ then from Eqs. (\ref{eq1}) and (\ref{eq2}) one obtains two flat band solutions which are tangent to the dispersive bands at $\Gamma$:
\begin{equation}
E_{1,4}=\pm \frac{3|t|}{2} \; .
\end{equation}
\begin{figure}[h!]
\onecolumngrid
	\centering
	\includegraphics[width=7cm]{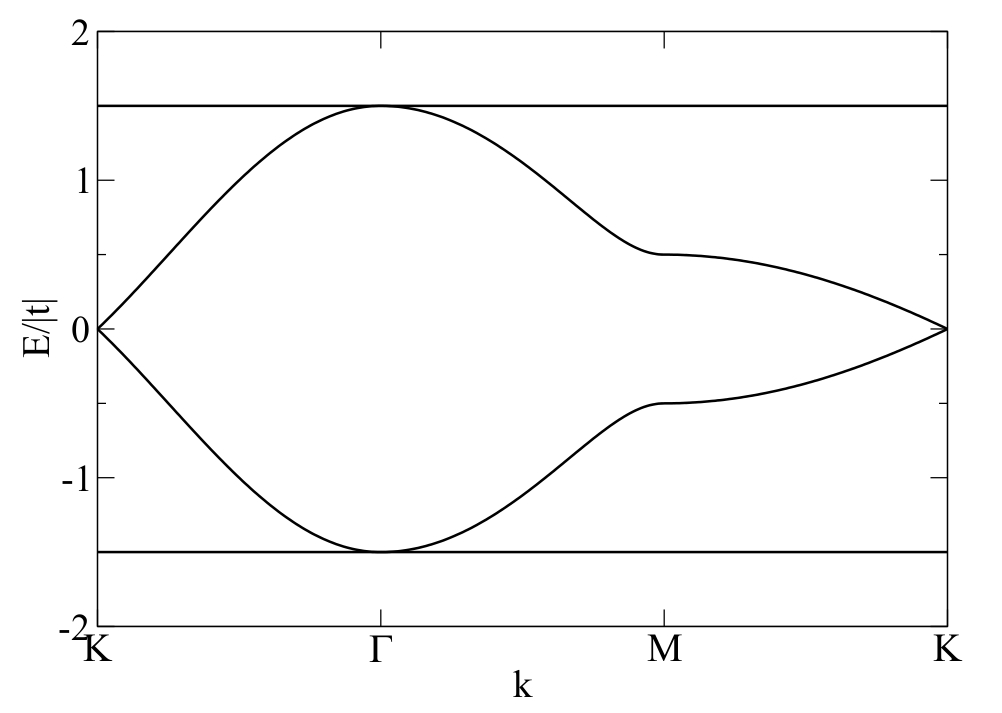}
	\caption{\label{fig:graphene-px-py-flat-band} $p_x-p_y$ graphene band structure}
\end{figure}
The localized eigenstates verify $\sum_{j } (\bm{\hat{n}}_{ij} .\bm{C}_{j} )=0 $ which is equivalent to the condition obtained for the simple Kagome lattice ($ \forall i, \;C_i=0$) except that now it is a vectorial equation and there are two solutions corresponding to the lowest and highest band. One can then build localized states on hexagonal rings as shown in Fig. \ref{fig:graphene-px-py-localized-states}. 

Let us now build the periodic solutions (Bloch states) of the flat band eigenstates. Starting from  periodic solutions $\bm{C}_i=e^{i\bm{k}.\bm{R}_i} \bm{C}_{A(B)}(\bm{k})$ and inserting into Eq.(\ref{eq1}) (or Eq.(\ref{eq2}))  one obtains:
\begin{equation}
\label{eq:flat-band-rel}
\bm{C}_A.\bm{\hat{n}}=\pm e^{i\bm{k}.\bm{n}} \bm{C}_{B}.\bm{\hat{n}}\;,
\end{equation}
and summing over the three possible vectors  $\bm{\hat{n}}$  ( $\bm{\hat{n}}_i$, $i=1,2,3$) in Eq.(\ref{eq:flat-band-rel}) it comes that $\bm{C}_B(\bm{k})$ is orthogonal to vector $\bm{u}(\bm{k})$ (Eq.\ (\ref{eq:def-u})) and therefore proportional to $\bm{u}^+-\bm{u}^-$ and similarly $\bm{C}_A(\bm{k})$ is proportional to $(\bm{u}^+-\bm{u}^-)^{\star}$. Close to $\Gamma$ ($\bm{k}\rightarrow 0$), $\bm{C}_B\approx \pm \bm{C}_A \approx \sum_i [\bm{k}.(\bm{n}_{i+1}-\bm{n}_{i-1})]\bm{\hat{n}}_i$ and these vectors are orthogonal to the wave vector $\bm{k}$. The flat modes are therefore transversal in the long wavelength limit. In the proximity of $K$ the three vectors $\bm{u}$, $\bm{u}^+$ and $\bm{u}^-$ are proportional to the vector of components $(1,\pm i)$ indicating opposite circular polarizations for the two lattices ($A$ and $B$).\\

The periodic solutions of the dispersive eigenstates are obtained in a similar fashion starting from Eqs.\ (\ref{eq1}) and Eq.\ (\ref{eq2}) for the the eigenvalues $E_{2,3}$ one obtains for any vector $\hat{ \bm{n} }$  connecting a $A$ site to a $B$ site.
\begin{subequations}
\begin{align}
\pm |\gamma(\bm{k})| \bm{C}_{A}(\bm{k}).\hat{ \bm{n} }  & + 3  \bm{C}_{B}(\bm{k}).\bm{\hat{n}} e^{i\bm{k}.\bm{n}} =b_A=e^{i\theta_{\bm{k}}/2}  \label{eq1bis}  \\
  3 \bm{C}_{A}(\bm{k}).\bm{\hat{n}} e^{-i\bm{k}.\bm{n}}  & \pm   |\gamma(\bm{k})| \bm{C}_{B}.\bm{\hat{n}}= -b_B =\mp e^{-i\theta_{\bm{k}}/2}    \; .
\label{eq2bis}
\end{align}
\end{subequations}
Summing over the connecting vectors  $\hat{ \bm{n} }$ we find that $ \bm{C}_{A}(\bm{k})$ is proportional to $\bm{u}(\bm{k}) e^{-i\theta_{\bm{k}}/2}$ and $\bm{C}_{B}(\bm{k})$  to $\mp e^{i\theta_{\bm{k}}/2}  \bm{u}^{\star}(\bm{k})$.

Finally it is interesting to note the striking similarity between the $p_x$-$p_y$ model and the simplified phonon model of graphene used in Ref.[\onlinecite{Guinea1981,CastroNeto2007}]. 

\subsection{Honeycomb-Kagome band structure}
\label{honeycomb-kagome-band}

For the honeycomb-Kagome the same procedure is applied {\sl i.e.} combining  ${\bm k}$ space and real space expansion of Schr\"{o}dinger equation. In our TB model there are no direct  Kagome-Kagome or honeycomb-honeycomb hopping and the Hamiltonian $H({\bm k})$ in $\bm k$ space takes the form:

\begin{equation}
H({\bm k})=
t\begin{pmatrix}
  0               &    0     &   0  &   e^{i{\bm k}.{\bm \tau}_1} &   e^{-i{\bm k}.{\bm \tau}_1}\\
  0               &    0     &   0  &   e^{i{\bm k}.{\bm \tau}_2} &   e^{-i{\bm k}.{\bm \tau}_2}\\
  0               &    0     &   0  &   e^{i{\bm k}.{\bm \tau}_3} &   e^{-i{\bm k}.{\bm \tau}_3}\\
  e^{-i{\bm k}.{\bm \tau}_1} &   e^{-i{\bm k}.{\bm \tau}_2} &  e^{-i{\bm k}.{\bm \tau}_3} &  0 & 0 \\
  e^{i{\bm k}.{\bm \tau}_1} &   e^{i{\bm k}.{\bm \tau}_2}  &  e^{i{\bm k}.{\bm \tau}_3}   &  0 & 0 
 \end{pmatrix}  \; .
\end{equation}
Where  ${\bm \tau}_i=\pm{\bm n_i}/2$   are the connecting vectors  (see Fig.\ \ref{fig:honeycomb-kagome}).
Taking the square of the Hamiltonian\cite{closed-path}:

\begin{equation}
H^2({\bm k})=
t^2\begin{pmatrix}
  2                &    2 \cos k_2     &   2 \cos k_1  &  0 &   0\\
 2 \cos k_2   &    2     &    2 \cos k_3  &  0 &  0\\
  2 \cos k_1 &  2 \cos k_3 &   2  &  0 &  0 \\
 0 &  0  & 0 &  3 & \gamma({\bm k}) \\
 0 &  0  & 0 &  \gamma^{\star}({\bm k}) & 3 \\
 \end{pmatrix}  \;,
\end{equation}
we recover a block diagonal matrix where the two block diagonal matrices have the form of a pure Kagome and honeycomb hamiltonian, respectively, that can be easily diagonalized. The energy dispersion can  finally be recast in the compact form:

\begin{equation}
E_{1,2,4,5}({\bm k})= \pm t  \sqrt{3\pm |\gamma({\bm k})|}  \quad  ; \quad  E_{3}=0  \; .
\end{equation}
\begin{figure}[h]
\onecolumngrid
	\centering
	\includegraphics[width=7cm]{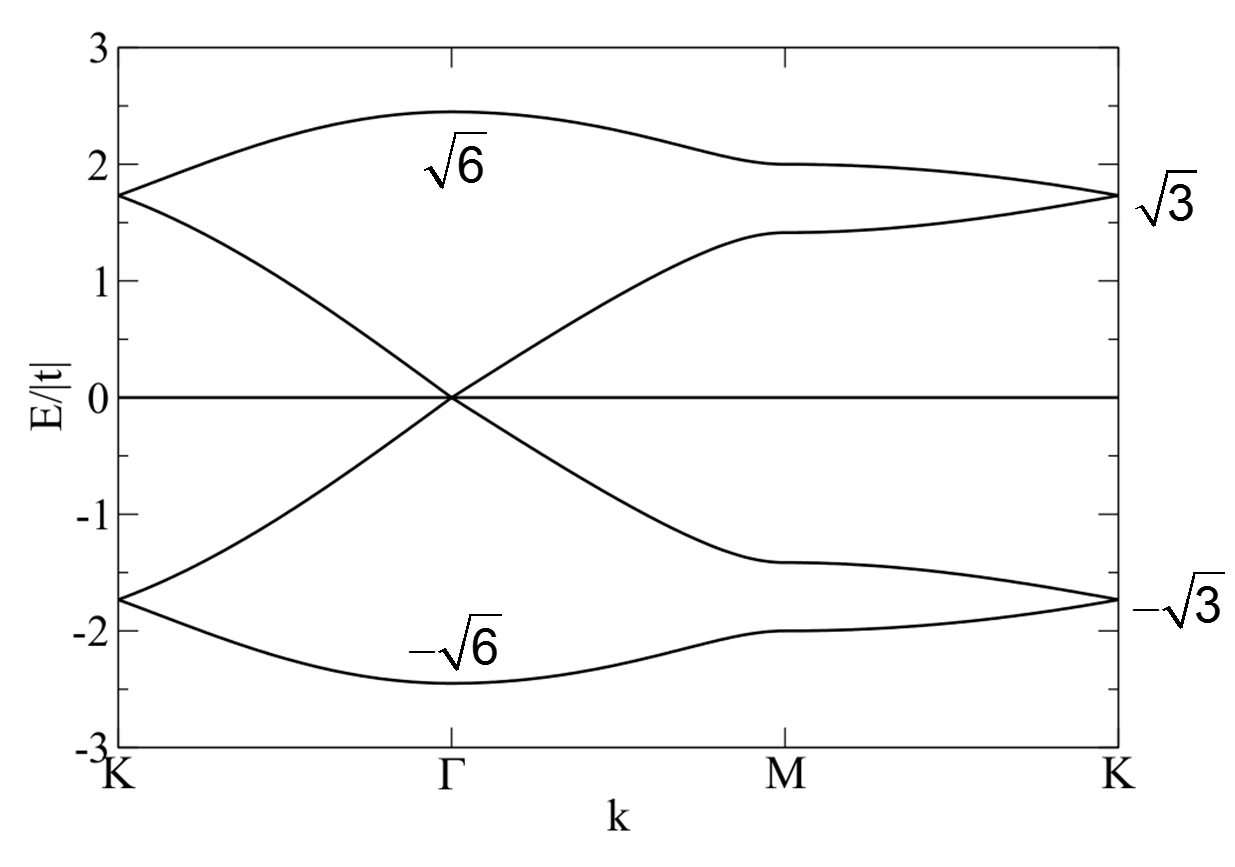}
	\caption{\label{fig:honeycomb-kagome-band} Honeycomb-Kagome band structure with the same on-site energy on honeycomb and kagome sites.  }
\end{figure}
There is a flat band at $E=0$ of states localized on the Kagome lattice and two linearly dispersive bands in the vicinity of the $K$ points at energies $\pm t \sqrt{3}$.  An additional linearly dispersive band appears in the center of the Brillouin zone  where the flat band crosses at zero  energy (see Fig.\ \ref{fig:honeycomb-kagome-band}). The existence of this linearly dispersive band only exists in the very specific case where all the atoms of the unit cell are equivalent. If on-site levels $\Delta$ are added to the TB Hamiltonian on the atoms of the Kagome lattice  the dispersion relation becomes:

\begin{figure}[htbp]
\subfloat[ $\Delta>0$]{\includegraphics[width=5cm]{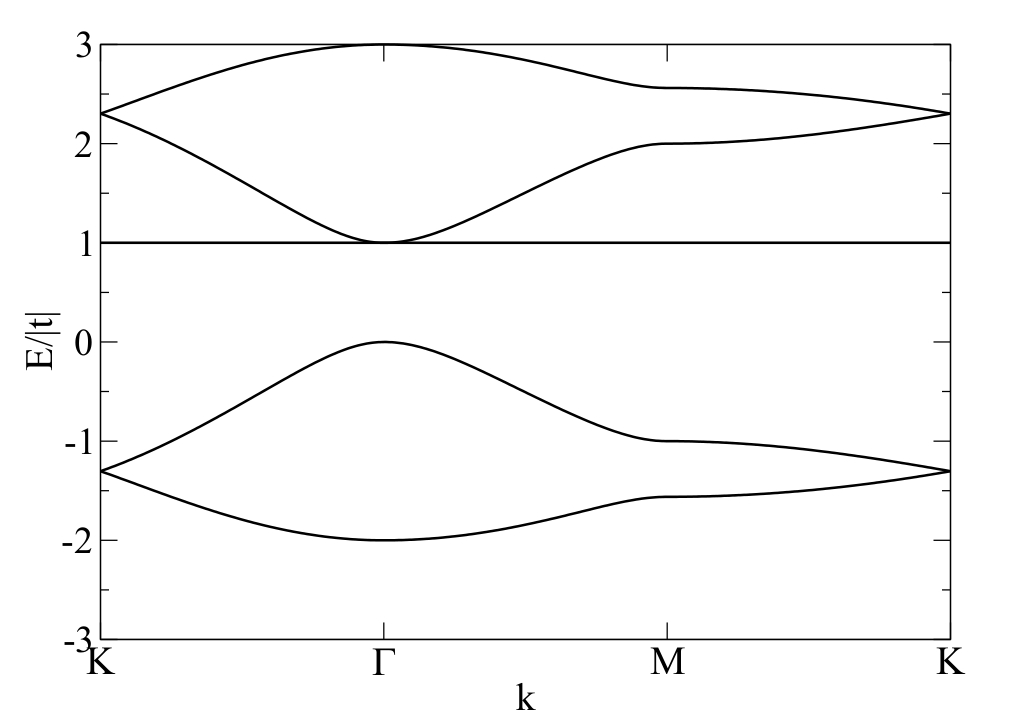}}
\subfloat[$\Delta<0$]{\includegraphics[width=5cm]{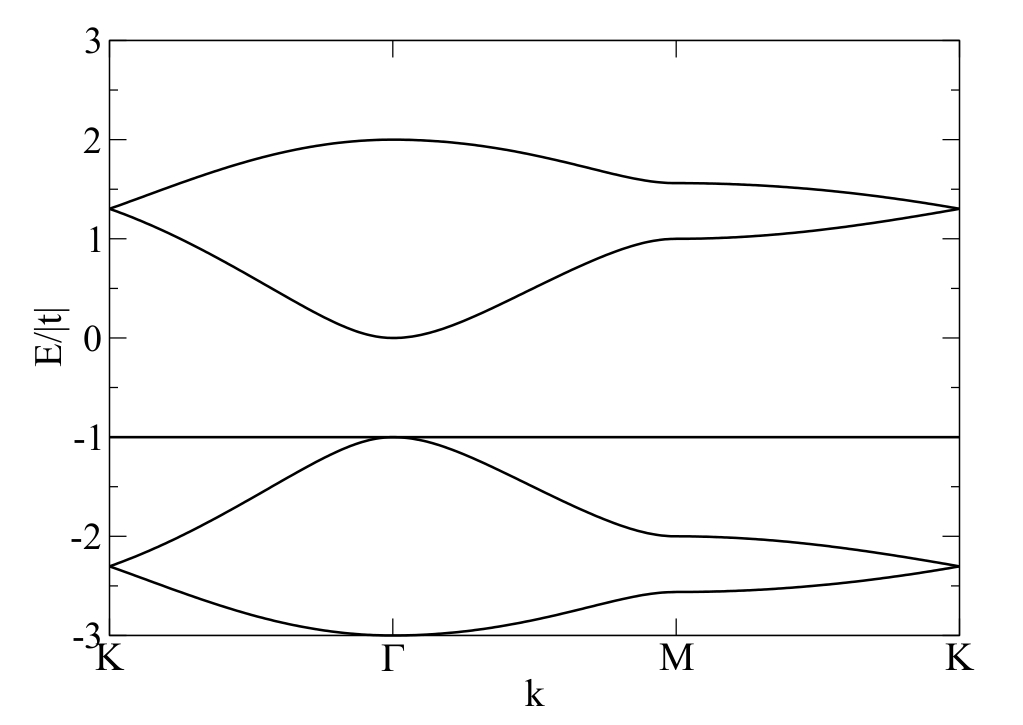}}
\subfloat[$\Delta<0$]{\includegraphics[width=5cm]{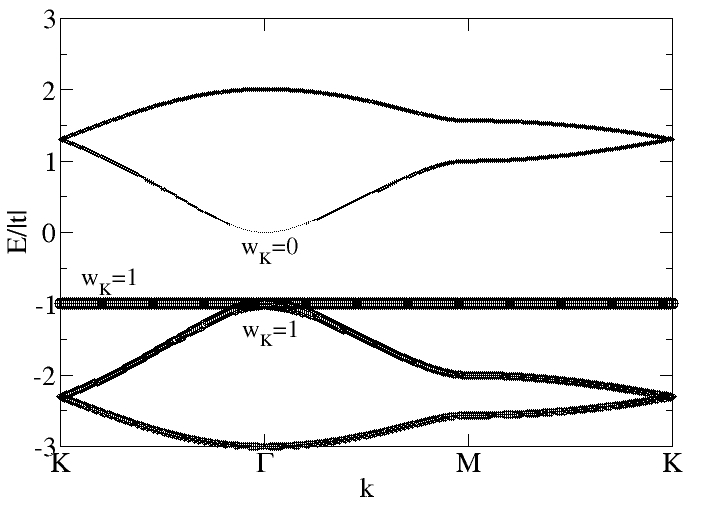}}
	\caption{\label{fig:honeycomb-kagome-band-oniste} Honeycomb-Kagome band structure with a) positive  and b) negative onsite $\Delta=\pm t$ on the Kagome lattice. c) is the same case as b) but where the size of the circles is proportional to the component of the eigenfunction on the  Kagome sites $w_K$. It appears that the flat band states are uniquely localized on the Kagome sites. At the $\Gamma$ point the eigenstate of the dispersive band  touching the flat band are also localized on the kagome lattice while the eigenstate of the upper band at its minimum is on the honeycomb lattice.}
\end{figure}

\begin{equation}
\label{eq:honeycomb-kagome}
E_{1,2,4,5}({\bm k})=  \frac{\Delta}{2}  \pm  \sqrt{   \frac{\Delta^2}{4}+t^2(3\pm |\gamma({\bm k})|)     }   \quad  ; \quad  E_{3}=\Delta  \; .
\end{equation}
A direct band gap $\Delta$  is opened at the center of the zone but the two other linearly dispersive bands in $K$ remain. The flat band $E_{3}=\Delta$ (of states localized on the Kagome lattice) touches the upper or lower dispersive bands at $\Gamma$ depending on the sign of the on-site levels as illustrated in Fig.\ \ref{fig:honeycomb-kagome-band-oniste}a and \ref{fig:honeycomb-kagome-band-oniste}b. Such  dispersion relation features have been described recently in the context of molecular graphene produced by adsorbed CO molecules on a copper (111)\cite{Paavilainen2016}.

Once again it is possible to derive the energy spectrum from an expansion of Schr\"{o}dinger equation in real space. Starting from site A or 1,respectively, one gets $E\, c_A=t\,(c_1+c_2+c_3)=t\,C_A$ and
 $E\, c_1=t\,(c_A+c_B)$. Combining both equations it comes $E^2 c_1=t^2(C_A+C_B)$. Similar equations are obtained for sites 2 and 3, and summing up the three equations one gets:
 $(E^2-3t^2)\,C_A=t^2(C_B+C_{B_1}+C_{B_2})$. By analogy with the equation obtained for graphene one recovers the dispersive bands $E= \pm t  \sqrt{3\pm |\gamma({\bm k})|} $. The flat band is obtained in the case where for all $i,C_i=0$, which gives $E=0$. The flat bands localized states are of the same nature as the one of the pure Kagome lattice.

\subsection{px-py honeycomb s-Kagome band structure}
\label{pxpy-honeycomb-s-kagome-band}

\begin{figure}[h]
\onecolumngrid
	\centering
	\includegraphics[width=4cm]{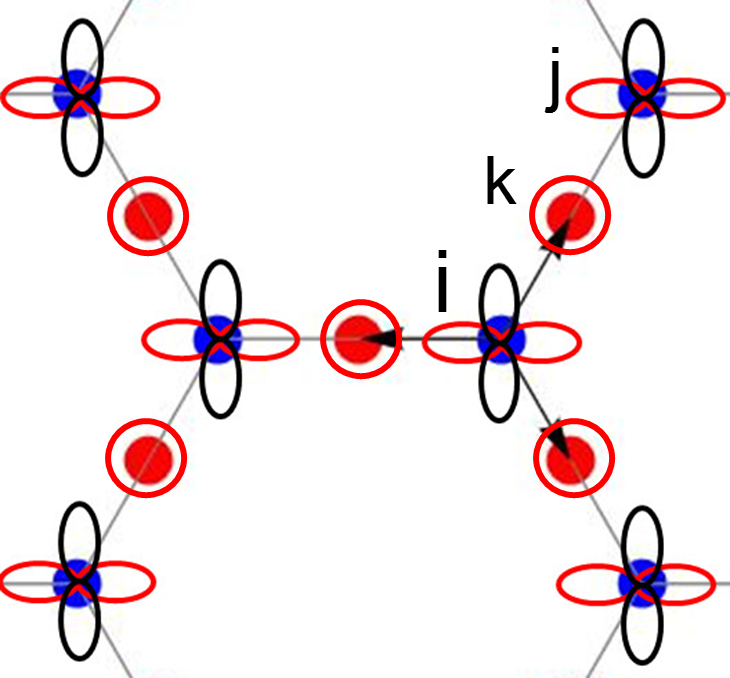}
	\caption{\label{fig:pxpy-honeycomb-s-kagome-lattice} px-py-honeycomb-s-Kagome lattice}
\end{figure}

\begin{figure}[h]
\onecolumngrid
	\centering
\includegraphics[width=10cm]{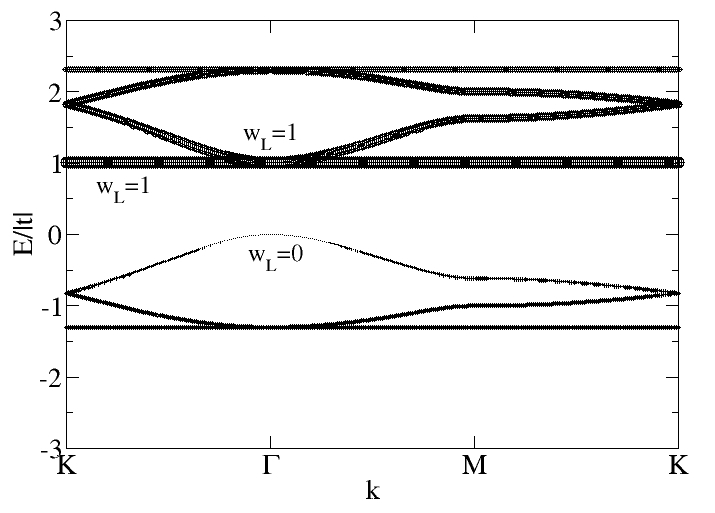}
\caption{\label{fig:pxpy-honeycomb-s-kagome-band}Electronic band structure of a $p_xp_y$ honeycomb $s$ kagome lattice with an on-site $\Delta=|t|$.The size of the circles is proportional to the component of the eigenfunction on the  honeycomb  sites $w_L$. The flat band $E=\Delta$ is a pure honeycomb state ({\sl i.e.} $w_L=1$) while the two other flat band  ($E=\frac{\Delta\pm\sqrt{\Delta^2+12t^2}}{2}$.) states have components on both lattices.}
\end{figure}

Let us us now  consider the case where the honeycomb sites are occupied by $p_x-p_y$ orbitals and the kagome sites by $s$ orbitals. Although no real materials can be directly described by such model we will see that it is very relevant in the case of the ligand-decorated honeycomb Kagome lattice.  The hopping integral between an orbital $p_{\lambda}$ of the honeycomb site $i$ and a $s$ orbital on the neigbhoring Kagome site $k$  is $t_{ik}^{\lambda}=t(\bm{\hat{\lambda}}.\bm{\hat{n}_{ik}})$.  Since the Kagome sites are in the middle of two honeycomb sites the normalized vector $\bm{\hat{n}_{ik}}$ is the same as the one connecting the two adjacent honeycomb sites $i$ and $j$ and therefore 
 $\bm{\hat{n}_{ik}}=\bm{\hat{n}_{ij}}$ (see Fig.\ \ref{fig:pxpy-honeycomb-s-kagome-lattice}).
Writing  Schr\"{o}dinger equation in real space one gets two equations relating the expansion coefficients $c_{i \lambda}$ and $c_k$ of the eigenfunctions on site $i$ (and orbital $\lambda$) and $k$:

\begin{subequations}
\begin{align}
(E-\Delta) c_{i \lambda}&=t \sum_k  (\bm{\hat{\lambda}}.\bm{\hat{n}}_{ik})c_k \label{eq:pxpy-honeycomb-s-kagome1}  \\
 Ec_k&=-t \sum_{j \lambda} (\bm{\hat{\lambda}}.\bm{\hat{n}_{kj}} )c_{j \lambda}\; ,   
 \label{eq:pxpy-honeycomb-s-kagome2}
\end{align}
\end{subequations}
where $\Delta$ is the on-site energy on the honeycomb lattice. Combining both equations and introducing the vector $\bm{C}_i=\sum_{\lambda} c_{i \lambda} \bm{\hat{\lambda}}$ one obtains:
\begin{equation}
E(E-\Delta)\bm{C}_i=t^2 \sum_j (\bm{\hat{n}}_{ij}.(\bm{C}_i-\bm{C}_j))\bm{\hat{n}}_{ij}  \; .
\end{equation}
Apart from the additional on-site levels this equation is very similar to Eq.\ (\ref{eq:pxpx-honeycomb}). Proceeding in a similar way we obtain four dispersive bands which expression is the same as the one of honeycomb-kagome (Eq. \ref{eq:honeycomb-kagome}), save and except for a renormalization of the hopping integral $t\rightarrow t/\sqrt{2}$:
\begin{equation}
E=\frac{\Delta\pm\sqrt{\Delta^2+2t^2(3\pm|\gamma({\bm k})|)}}{2} \; .
\end{equation}
The band structure shows a gap at $\Gamma$ equal to $\Delta$.
There are also three flat bands tangent at $\Gamma$ to the dispersive bands at energies:
\begin{equation}
E=\frac{\Delta\pm\sqrt{\Delta^2+12t^2}}{2} \quad \text{and} \quad E=\Delta \; .
\end{equation}
There is no flat band at $E=0$ since this would lead to a null eigenfunction. The flat band at the energy of the honeycomb on-site $\Delta$ corresponds to a purely honeycomb state with no weight on the Kagome sites while the two other flat bands have components on both honeycomb and kagome lattices. The band structure is shown in Fig.\ \ref{fig:pxpy-honeycomb-s-kagome-band} for a positive  on-site level equal to the hopping integral $\Delta=|t|$. 

\subsection{$\alpha$-graphyne band structure}
 \label{graphyne-band}
 
$\alpha$- graphyne is an example of honeycomb-Kagome lattice but with dimers of ``Kagome atoms'' in between two first neigbour atoms of the honeycomb lattice. Although it has not been synthetized yet, small molecular units have been obtained and inserted between two gold electrodes to form molecular junctions\cite{ZhihaiL2015}. The electronic structure of various forms of $\alpha$-graphyne have also been presented in the literature \cite{Malko2012,Zheng2013,Sun2015,Kang2015}. For the sake of generality we consider two hopping integrals $t$ and $t'$ connecting two Kagome atoms or a Kagome with a honeycomb atom, respectively (see Fig. \ref{fig:graphyne}). The derivation of the energy dispersion is much more straightforward in real space. Starting from site $i=1,2,3$ or $i=1',2',3'$ one gets $E c_1=t'c_A+tc'_1$ and $E c'_1=t'c_B+tc_1$, respectively, while for site $A$ and $B$ we have $E c_A=t'(c_1+c_2+c_3)=t'C_A$ and  $E c_B=t'C_B$. Combining these equations one can easily obtain the relation $E(E^2-3t'^2-t^2)C_A=tt'^2(C_B+C_{B_1}+C_{B_2})$ where we recognize an equation similar to the one of graphene. The dispersion relation is therefore expressed as a solution of a third degree equation:
\begin{equation}
\label{eq:graphyne-band}
E(E^2-3t'^2-t^2)=\pm tt'^2 |\gamma({\bm k})|  \; ,
\end{equation}
and the flat bands are obtained whenever $\forall i \, \,C_i=0 $ which gives two non dispersive bands $E=\pm t$ whatever the value of $t'=\alpha t$ (where $\alpha$ is a dimensionless factor). 
It can be shown using Cardano's method that the cubic equation (\ref{eq:graphyne-band}) has three real solutions that can be expressed analytically:
\begin{equation}
\label{eq:graphyne-band-cardano}
E_1= \pm t \alpha^{\frac{2}{3}}(v+\bar{v}) \quad ; \quad  E_2= \pm t \alpha^{\frac{2}{3}}(j v+ \bar{j} \bar{v})  \quad ; \quad  E_3= \pm t \alpha^{\frac{2}{3}}(\bar{j} v+ j \bar{v}) \; ,
\end{equation}
with $v=\sqrt[3]{\frac{|\gamma|+i\sqrt{\delta}}{2}}$  where $\delta=\frac{4}{27}\frac{(3 \alpha^2+1)^3}{\alpha^4}-|\gamma|^2$ is always a positive number.\\

\begin{figure}[htbp]
\subfloat[ $t'=t/2$]{\includegraphics[width=5.5cm]{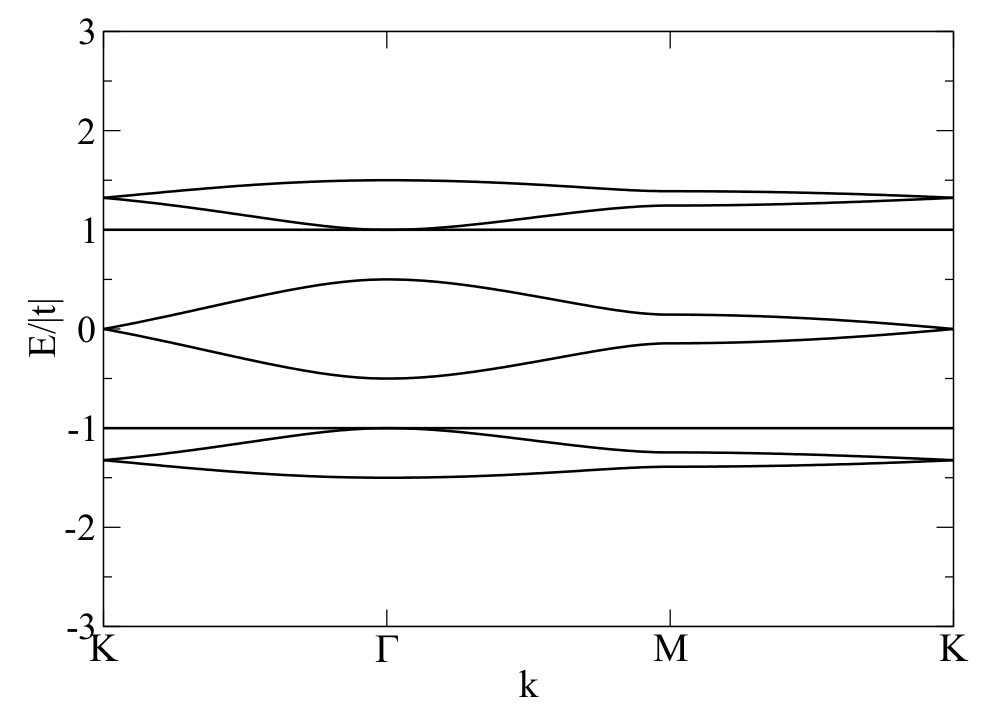} }
\subfloat[ $t'=\sqrt{2/3}t$]{\includegraphics[width=5.5cm]{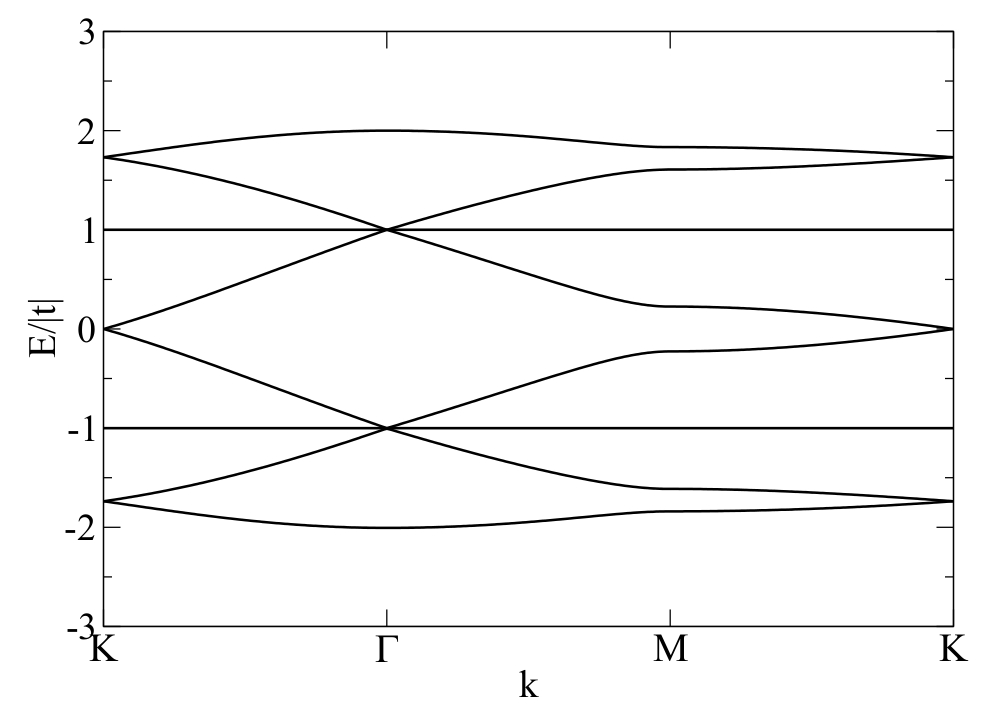} }
\subfloat[$t'=t$]{\includegraphics[width=5.5cm]{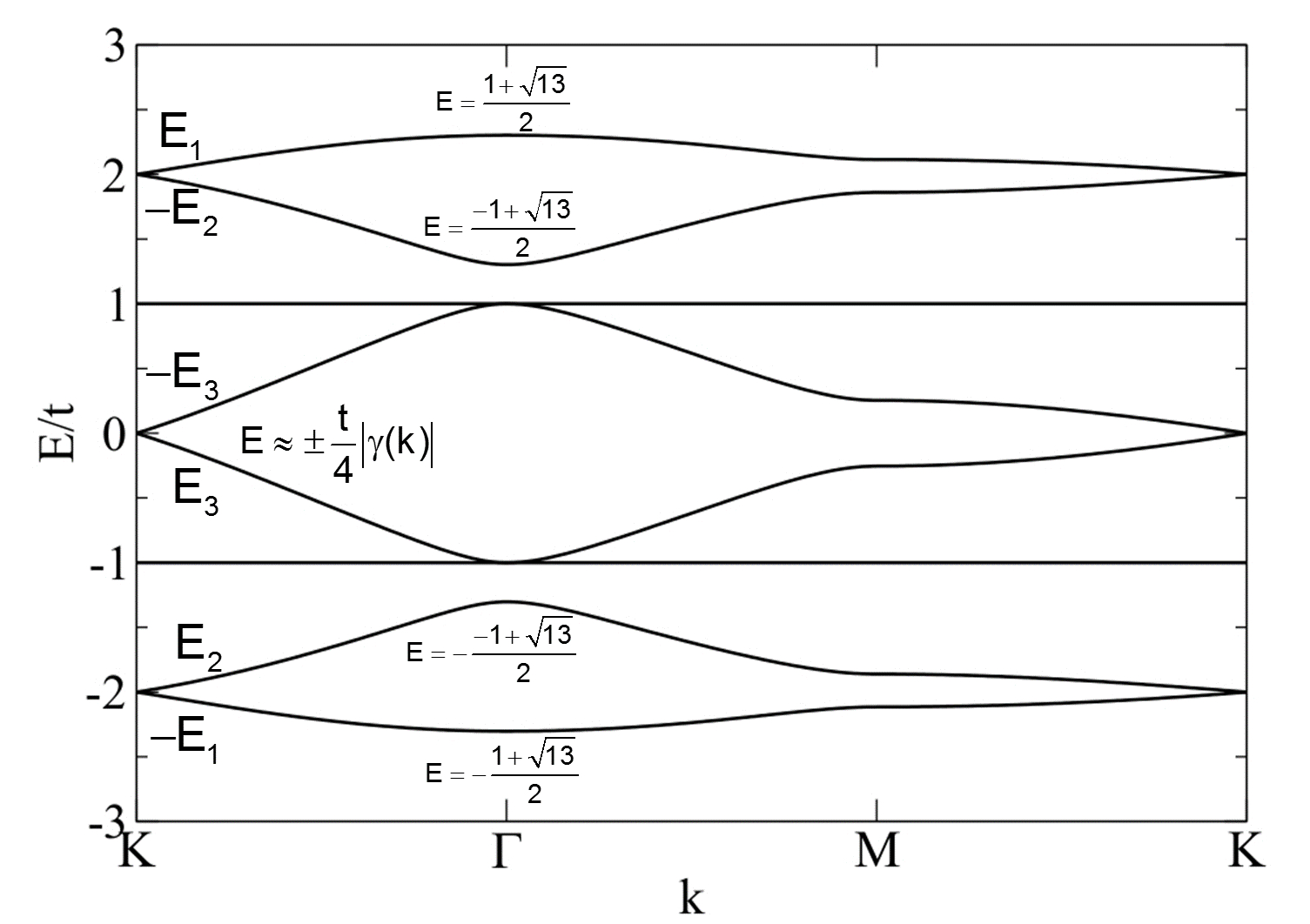} }
	\caption{\label{fig:graphyne-band}$\alpha$- graphyne  band structure for a)  $t'=t/2$  , b)  $t'=\sqrt{2/3}t$  and c) $t'=t$ }
\end{figure}

The band structure of $\alpha$-graphyne is plotted in Fig.\ \ref{fig:graphyne-band} (note that the spectrum is symmetric\cite{closed-path}) in the three cases $t'=t/2$, $t'=\sqrt{2/3}t$ and $t'=t$. The band structure is composed of three separate sets of dispersive bands: An upper and lower set which are opposite to each other and a middle set centered on zero energy. Each set of dispersive band shows a linear dispersion at $K$ ($\gamma(K)=0$) at energies $E=\pm t \sqrt{3\alpha^2+1}$ and $E=0$. The flat bands at $E=\pm t$ are always touching the dispersive bands at the center of the zone. However one observes a change of morphology at $t'=\sqrt{2/3}t$ since below this critical value the flat bands are touching the upper and lower sets while above this critical value the flat bands are touching the middle set. For $t'=\sqrt{2/3}t$  $\delta(K)=0$ so that the upper and lower sets are touching the middle set and an  accidental linearly dispersive band appears at the center $\Gamma$  of the zone. For the sake of generality (although it is not relevant in the case of Carbon $\alpha$-graphyne) it is still interesting to consider the case where $t'$ is large compared to $t$, {\sl i.e.} $\alpha>>1$. By developing  Eq.\ (\ref{eq:graphyne-band-cardano}) with respect to $\alpha$, one can show that the upper and lower dispersive bands are shifted linearly upwards and downwards and reach a stable dispersion relation. The middle set reaches as well a stable dispersion relation:

\begin{equation}
E_{\text {upper}}^{\alpha>>1}= -\sqrt{3\alpha^2+1}\pm \frac{ |\gamma|}{6} \quad ; \quad E_{\text{lower}}^{\alpha>>1}= -E_{\text{upper}}^{\alpha>>1} \quad ; \quad E_{\text{middle}}^{\alpha>>1}= \pm \frac{ |\gamma|}{3}  \; .
\end{equation}

\subsection{Ligand decorated honeycomb-Kagome band structure}
\label{decorated-honeycomb-kagome-band}

\begin{figure}[h]
\onecolumngrid
	\centering
	\includegraphics[width=8cm]{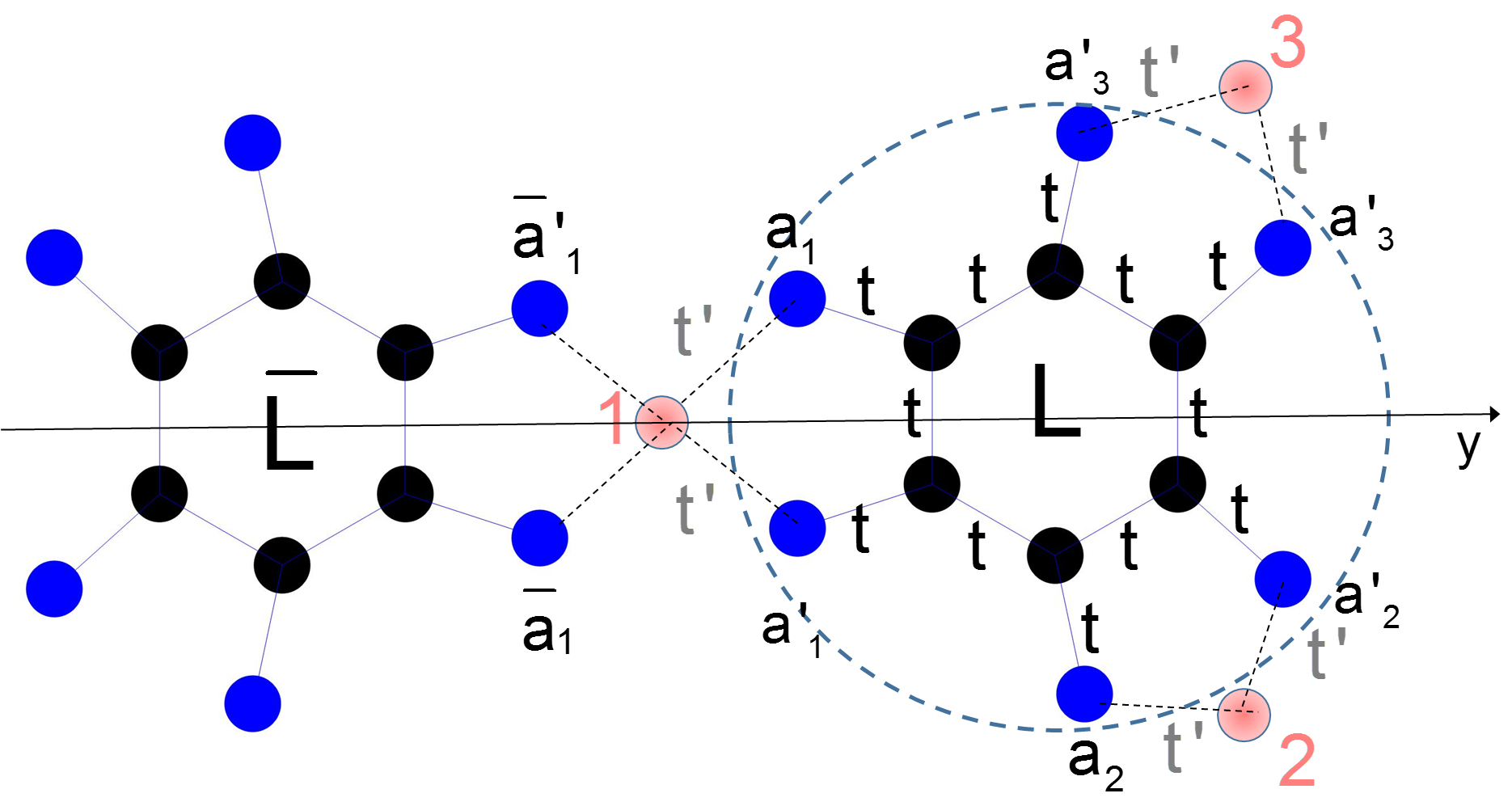}
	\caption{\label{fig:ligand-t} Schematic representation of the unit-cell of  the ligand-decorated honeycomb kagome lattice where we have indicated the various sites and notations used in the main text. The ligand described in the Figure (delimited by a dahsed circle) is the one that we have used for the numerical applications.}
\end{figure}

Let us now consider the band structure of a very general class of lattices: the ligand-decorated honeycomb Kagome class. The honeycomb lattice sites are occupied by ligands L with $C_{3v}$ symmetry while the Kagome lattice sites are occupied by single atoms (or ions) hearafter denoted Kagome atoms. Each Kagome atom is connected to two ligands $ L$ and its inversion symmetric $\bar{L}$ via two atoms of the ligand (see Fig.\ \ref{fig:ligand-t} where for example atom 1 is connected to $L$  ($\bar{L}$) via sites $a_1$ and $a'_1$ ($\bar{a}_1$ and $\bar{a}'_1$)).  In the following we consider a simple ligand made of an hexagonal (benzene-like) ring where each atom (in black) of the ring is also connected to a single (blue) atom. The hopping integrals between the first neighbours of the ligand is taken equal to $t$ while the kagome sites will be connected to a given ligand via two blue atoms with hopping integral $t'$ as illustrated in Fig.\ \ref{fig:ligand-t}. However, our demonstration below applies to the general case of any ligand of  $C_{3v}$ symmetry.
The Hamiltonian of the system can naturally be split into a Kagome Hamiltonian $H_K^0$ and a ``ligand'' Hamiltonian $H_L^0$ that are connected via hopping elements $V_{LK}$, and can formally be written:
\begin{equation}
H=
\begin{pmatrix}
H_K^0    &  V_{KL} \\
 V_{LK}   &  H_L^0
\end{pmatrix}  \;  .
\end{equation}
Splitting as well the components  of the eigenstates $|\Phi\rangle$  into Ligand  $|\Phi_K\rangle$ and Kagome $|\Phi_L\rangle$  components: $|\Phi_{K} \rangle=\sum_{i \in K } c_i |i\rangle$  and $|\Phi_{ L} \rangle=\sum_{i \in L} c_i |i\rangle$, Schr\"{o}dinger equation reads:
\begin{subequations}
\begin{align}
E |\Phi_{K} \rangle & = H_K^0 |\Phi_{K} \rangle+V_{KL}|\Phi_{K} \rangle   \label{eqShroK}  \\
E |\Phi_{L} \rangle & = H_L^0 |\Phi_{L} \rangle+V_{LK}|\Phi_{L} \rangle   \label{eqShroL} \; .
\end{align}  
\end{subequations} 
Let $G_K$ and $G_L$ be the Green functions in subspaces $K$ and $L$ it comes that:
\begin{subequations}
\begin{align}
G_{K}(z)  & =( z-H_K^0-V_{KL}G_L^0 V_{LK})^{-1}=(z-H_K)^{-1} \label{eqGreenK}  \\
G_{L}(z)  & =( z-H_L^0-V_{LK}G_K^0 V_{KL})^{-1} =(z-H_L)^{-1}  \label{eqGreenL} \;.
\end{align}
\end{subequations} 
where $G_{L}^0=(E+i\epsilon -H_{L}^0)^{-1}$  and $G_{K}^0=(E+i\epsilon -H_{K}^0)^{-1}$ are the Green function matrices of the isolated ligand and Kagome respectively. The poles of the Green functions are eigenvalues of the system. If the states are coupled (which is the general case) one can look for poles of $G_K$. We will see that this gives us most of the eigenstates except some that are decoupled from the Kagome sites and that appear only as poles of $G_L$. Let $c_i=\langle i  |\Phi_{K} \rangle$ be the component of the eigenfunctions $ |\Phi_{K} \rangle$ on site $i$. Projecting the Schr\"{o}dinger equation $H_K |\Phi_{K} \rangle=E |\Phi_{K} \rangle$ on site $1$, one gets:
\begin{equation}
E c_1 =\langle 1| H_K^0+V_{KL}G_L^0V_{LK}|\Phi_{K} \rangle = \sum_i  c_i \langle 1|  H_K^0+V_{KL}G_L^0V_{LK}|i\rangle.
\end{equation}
Investigating the number of paths that leads to site 1 in three jumps {\sl i.e.} starting from a Kagome atom  $i$ jumping on ligand (L), then ``self jumping'' on ligand and  last jump towards atom 1  (see Fig.\ \ref{fig:ligand-t}) and taking into account the symmetry of the ligand one obtains:
\begin{align}
\label{eq:eq1}
Ec_1 &= 4t'^2 F_0 c_1 +2t'^2 F_1 (c_2+c_3+c_{\bar{2}}+c_{\bar{3}}) \\
 &= 4t'^2 (F_0-F_1) c_1 +2t'^2 F_1 (\underbrace{c_1+c_2+c_3}_{\text{\normalsize $C_A$}}+(\underbrace{c_1+c_{\bar{2}}+c_{\bar{3}}}_{\text{\normalsize $C_B$}} )  \nonumber ,
\end{align}
where $t'$ is the hopping integral between a Kagome atom and its nearest neighbour ligand and $2 F_0(E)=\langle a_1+ a'_1 | G_L^0 |  a_1+ a'_1 \rangle$ and  $2 F_1(E)=\langle a_1+ a'_1 | G_L^0 |  a_2+ a'_2 \rangle$. Applying the same procedure for site 2 and 3 and summing up it comes:
\begin{equation}
\frac{E}{4t'^2}C_A = F(E) C_A +\frac{1}{2}F_1(E) (3 C_A+C_B+C_{B_1}+C_{B_3}) ,
\end{equation}
where $F(E)$ is a diagonal element of the Green function  of the ligand $G_L^0$:
\begin{equation}
F(E)=F_0(E)-F_1(E)=\frac{1}{4} \langle a_1+ a'_1 -a_2 -a'_2| G_L^0 |  a_1+ a'_1 -a_2 -a'_2 \rangle.
\end{equation}
As usual the case where $ \forall i   \, \, C_i=0$ leads to the Kagome flat bands whose energy levels  (using Eq.\ \ref{eq:eq1}) are given by :

\begin{equation}
\label{eq:honeycomb-kagome-flat}
\frac{E}{4t'^2}=F(E)  \; ,
\end{equation}
while the dispersive bands are obtained once again by analogy with Eq. (\ref{eq:Shro-graphene-real}):
\begin{equation}
\label{eq:honeycomb-kagome-dispersion}
\frac{E}{4t'^2}=F(E)+\frac{1}{2} F_1(E)(3\pm |\gamma({\bm k})|) ,
\end{equation}
and then, the dispersion relation can formally be written as $E=f^{-1}(\pm |\gamma({\bm k})|)$ where the function $f^{-1}$ has several branches.\\

Let us now count the number of  flat and dispersive solutions.
The flat bands are obtained as the intersection between the straight line $E/4t'^2$ with the function $F(E)$ which is  a diagonal element of $G_L^0$.  The isolated  ligand Green function can be written in its eigenstates basis $|\alpha\rangle$:
\begin{equation}
G_ L^0=\sum_{\alpha}  |\alpha\rangle \frac{1}{E-E_{\alpha}}\langle \alpha |.
\end{equation} 
The ligand $L$ has the symmetry of a triangle  ($C_{3v}$) and there are three types of symmetry: A$_1$ (invariant), A$_2$ (alternate with respect to mirror symmetry) and the double degenerate one, E,  which transforms as the component of a bidimensional vector ($x$, $y$).
It is clear that for states of  symmetry A$_1$ and A$_2$ the components $ \langle \alpha |  a_1+ a'_1 -a_2 -a'_2 \rangle$  vanish since $\langle \alpha |  a_1\rangle=\langle \alpha |  a'_1\rangle=\langle \alpha |  a_2\rangle=\langle \alpha |  a'_2\rangle$ for A$_1$ states and $\langle \alpha |  a_1\rangle=-\langle \alpha |  a'_1\rangle$ and   $\langle \alpha |  a_2\rangle=-\langle \alpha |  a'_2\rangle$ for A$_2$ states. Therefore the E states are the only ones contributing to $F(E)$ and the equation to solve is of the type:
\begin{figure}[h]
\onecolumngrid
	\centering
	\includegraphics[width=9cm]{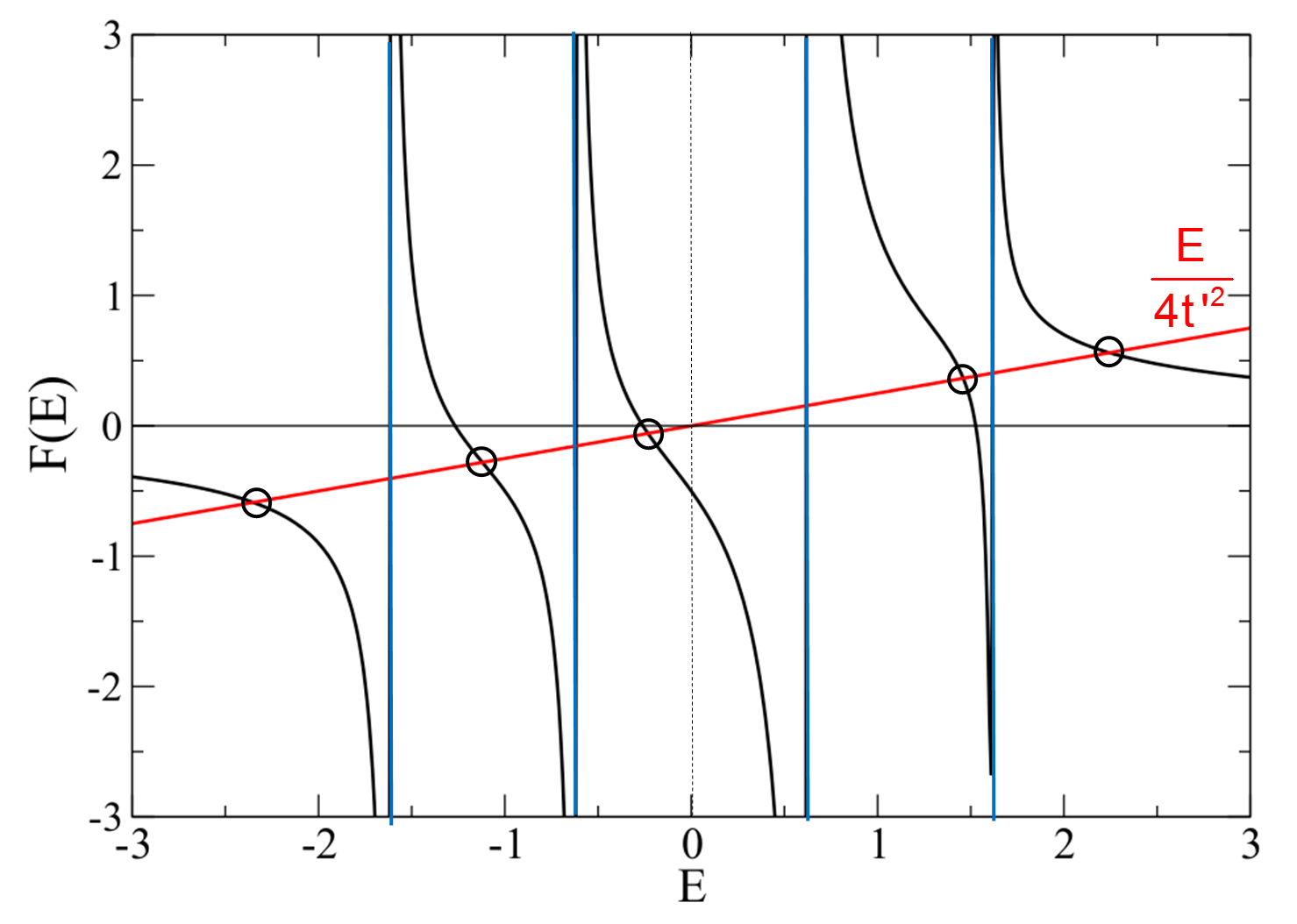}
	\caption{\label{fig:F-Green} Graphical representation of the function $F(E)$ and its intersection with the function $\frac{E}{4t'^2}$.  }
\end{figure}
\begin{equation}
\label{eq:honeycomb-kagome-flat2}
\frac{E}{4t'^2}=\sum_{\alpha \in E} \frac{\Lambda_{\alpha}}{E-E_{\alpha}},
\end{equation}
where $\Lambda_{\alpha}$ is a coefficient proportional to the weight of the E eigenstates on the sites $a_i$ connecting the ligand to the Kagome sites. The number of solutions of Eq.\ (\ref{eq:honeycomb-kagome-flat2}) is therefore equal to $n_E+1$, where $n_E$ is the number of states of E symmetry (without counting their double degeneracy). To understand how the Kagome flat bands evolve when the hopping integral $t'$ varies, it is convenient to solve Eq.\ (\ref{eq:honeycomb-kagome-flat2}) graphically as shown in Fig.\ \ref{fig:F-Green}. In particular when $t'/t \gg 1$ the lowest and largest solutions tends to $\pm \infty$ while the other are in between two eigenvalues of the E states. \\

From the dispersion relation  (\ref{eq:honeycomb-kagome-dispersion}) it appears that since $\gamma(0)=3$ the dispersive bands are touching the Kagome flat bands in $\Gamma$ ($\bm{k}=0$).The number of solutions of Eq.\ (\ref{eq:honeycomb-kagome-dispersion}) depends on the states that participate to the Green functions $F(E)$ and $F_1(E)$. For $F(E)$ only E states have non-zero contribution but for $F_1(E)$ the A$_1$ states also participate. A$_2$ states do not contribute since  $\langle \alpha |  a_1\rangle=-\langle \alpha |  a'_1\rangle$ and   $\langle \alpha |  a_2\rangle=-\langle \alpha |  a'_2\rangle$. Therefore there are then $2(n_E+n_{A_1}+1)$ dispersive states (factor 2 being due to the two branches $\pm |\gamma({\bm k})|$) and then, $3n_E+2n_{A_1}+3$ states from the analysis of $H_K$. Since there are two ligands and three Kagome atoms in the unit cell, one expects a total of $n_{\text{total}}=2(n_{A_1}+n_{A_2}+2n_E)+3$ states, and therefore $2n_{A_2}+n_E$ states are missing. These states are necessarily pure ligand states. A first set of states is easily found since for the symmetry A$_2$ we have for any pair of connecting atoms $\langle a_i+a'_i| \alpha \rangle=0$, therefore in the unit cell there are $2n_{A_2}$ such states.  $n_E$ states per unit cell are still missing. However their amplitudes on the different ligand is necessarily correlated; otherwise there would be $2n_E$ such states. Looking for solutions of (\ref{eqShroK}) and  (\ref{eqShroL}) with $|\Phi_{K} \rangle=0$ we have:
\begin{subequations}
\begin{align}
0 & =V_{KL}|\Phi_{K} \rangle   \label{eqShroK2}  \\
E |\Phi_{L} \rangle & = H_L^0 |\Phi_{L} \rangle   \label{eqShroL2} \; .
\end{align}
\end{subequations} 
The second equation shows that $|\Phi_{L} \rangle$ is a  linear combination of states localized in the ligand and projecting Eq.\ (\ref{eqShroK2}) on site $1$ (see Fig. \ref{fig:ligand-t}) one gets: 
\begin{equation}
\label{eq:ligand-states}
\braket{a_1+a'_1+\bar{a}_1+\bar{a}'_1}{\Phi_{L}}=0  \; .
\end{equation}
Apart from the trivial $2n_{A_2}$  states of symmetry $A_2$ there are $n_E$ doubly degenerated  E states. Let us start from a combination of such states:
\begin{equation}
|\Phi_{L}^{\nu} \rangle=\sum_{\ell,\lambda} \phi_{\ell}^{\nu,\lambda} |\ell, \nu, \lambda \rangle,
\end{equation}
where the sum runs over the ligands $\ell$ of the system and on the two components $\lambda=x,y$ of the $\nu=1,..,n_E$   states of E symmetry. In  the scalar product $\braket{a_1+a'_1}{\ell,\nu,\mu}$ only the symmetric component with respect to the mirror symmetry leaving the two neigbouring ligands  $\ell=L$ and $\ell'=\bar{L}$ invariant  will contribute for a given Kagome atom ($1$ in Fig. \ref{fig:ligand-t}). In fact $\braket{a_1+a'_1}{\ell,\nu,\mu}$ is proportional to $\bm{\hat{\lambda}}.\bm{\hat{n}}_{\ell, \ell'}$ where $\bm{\hat{n}}_{\ell, \ell'}$ is the vector connecting ligand $\ell$ and ligand $\ell'$. Therefore:
\begin{equation}
\braket{a_1+a'_1}{\ell,\nu,\mu} \propto \bm{\phi}_{\ell}^{\nu}.\bm{\hat{n}}_{\ell, \ell'} \quad \text{with}  \quad  \bm{\phi}_{\ell}^{\nu}=\sum_{\lambda}  \phi_{\ell}^{\nu,\lambda} \bm{\hat{\lambda}},
\end{equation}
and using Eq.\ (\ref{eq:ligand-states}) it comes that for any link between two neigbhouring ligands  $\ell=L$ and $\ell'=\bar{L}$  we have:
\begin{equation}
\bm{\hat{n}}_{\ell, \ell'}.(\bm{\phi}_{\ell}^{\nu}-\bm{\phi}_{\ell'}^{\nu})=0  \; .
\end{equation}
This equation is  similar to Eq.\ (\ref{eq1}) obtained for the $p_xp_y$ honeycomb model in the case $b_i=0$ corresponding to one of the flat bands $E=3t/2$ whose eigenstate is shown in Fig. \ref{fig:graphene-px-py-flat-band} (state E$4$). Therefore this leads to the missing $n_E$ flat bands.
To summarize there are $n_E+1$ Kagome-like flat bands, $2n_{A_2}$  ligand-type (flat bands) originating from $A_2$ states and $n_E$ flat bands originating from the E states. The remaining $2(n_E+n_{A_1}+1)$ bands are dispersive.\\

In order to obtain a more intuitive picture,  it is useful to use a slightly different approach. One can start by pre-diagonalizing the  Hamiltonian of the ligand. The total Hamiltonian then looks like
\begin{equation}
H=
\begin{pmatrix}
H_K^0    &  V'_{KL} \\
 V'_{LK}   & \sum_{\alpha}  \ket{\alpha} E_{\alpha}  \bra{\alpha}
\end{pmatrix}
\end{equation}
The orbitals being of A$_1$, A$_2$ or E symmetry it is possible to recast the original problem  into an honeycomb-Kagome problem where the honeycomb sites are occupied by a multi-orbital ($s$ (A$_1$), $p_x p_y$ (E) and $p_z$ (A$_2$)) atom. The problem cannot be solved analytically since the  orbitals  $s$ $p_x$ and $p_y$ will interact via the ``bridge'' of the Kagome atom.  The coupling with A$_2$ ($p_z$) states is however forbidden and this will lead to $n_{A_2}$ doubly degenerated ligand flat bands that do not interact at all with any other band. The $p_x$-$ p_y$ (E) orbitals will also give rise to $n_E$  ligand flat band (single degeneracy) tangent at $\Gamma$ to the dispersive bands. Being ligand states their energy is fixed at the level of the ligand and are independent of the hopping $t'$ like the A$_2$ states. However since they interact with the other states  they cannot cross any dispersive band and they ``act'' as an impassable barrier that will block the dispersive states. The other $n_E+1$ flat bands are Kagome bands whose energy depends on the hopping integral between the kagome atoms and the ligand. The remaining $2(n_E++n_{A_1}+1)$ dispersive states have a dispersion proportional to the hopping integral $V'_{KL}$ between the Kagome site and the ligand state:
\begin{equation}
V'_{KL}=t'(\alpha_{1}+\alpha'_{1})  \;,
\end{equation}
where $\alpha_{1}=\braket{a_1}{\alpha}$ and $\alpha'_{1}=\braket{a'_1}{\alpha}$ are the coefficients of the eigenstate $\ket{\alpha}$ projected on the sites $a_1$ and $a'_1$, respectively. These coefficients depend on the energy level $E_{\alpha}$ of the eigenstate. For A$_2$ states $\alpha_{1}+\alpha'_{1}=0$ and we recover the flat band. There are no exact expression for a general ligand but in the case of the ligand described in Fig.\ \ref{fig:ligand-t} the spectrum is given by $E_{\alpha}=\cos k_{\alpha}\pm \sqrt{1+\cos^2 k_{\alpha}}$ with $k_{\alpha}=\frac{2 \pi}{6} \alpha$ ($\alpha=-2,-1,0,1,2,3$). $\alpha=0$ corresponds to A$_1$ states ($E_{\alpha}=(1\pm\sqrt{2})t$),  $\alpha=3$ to A$_2$ states ($E_{\alpha}=(-1\pm\sqrt{2})t$) and the couples $\alpha=\pm 1$ $\alpha=\pm 2$ to the E states ($E_{\alpha}=(\pm1\pm\sqrt{5})t/2$). The coefficient of the wavefunction on the connecting atoms $a_1$  and $a'_1$ is proportional to $1/\sqrt{1+(E_{\alpha}/t)^2}$ and therefore the amplitude of the dispersion will be larger for the states close to the zero of energy (or more precisely to the reference energy of the ligand). \\

\begin{figure}[htbp]
\subfloat[$t'/t=0.5$]{\includegraphics[width=9cm]{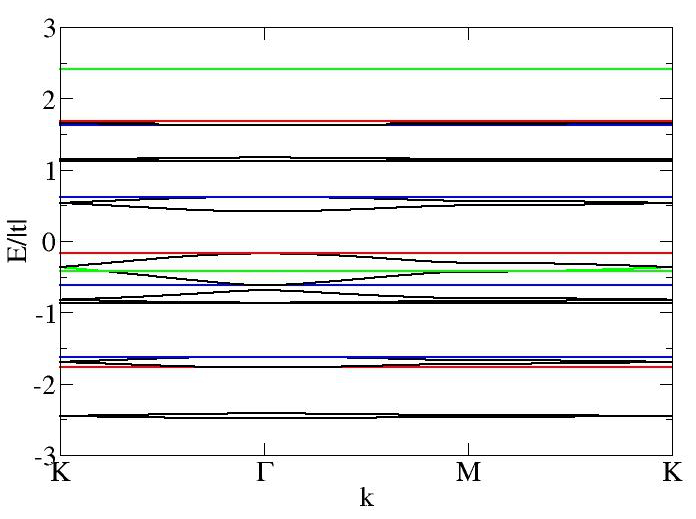} } \\
\subfloat[ $t'/t=1$]{\includegraphics[width=9cm]{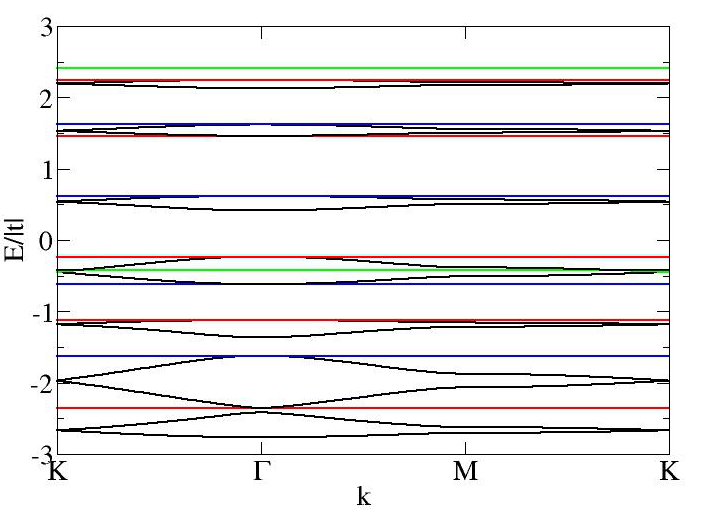}} \\
\subfloat[$t'/t=2$]{\includegraphics[width=9cm]{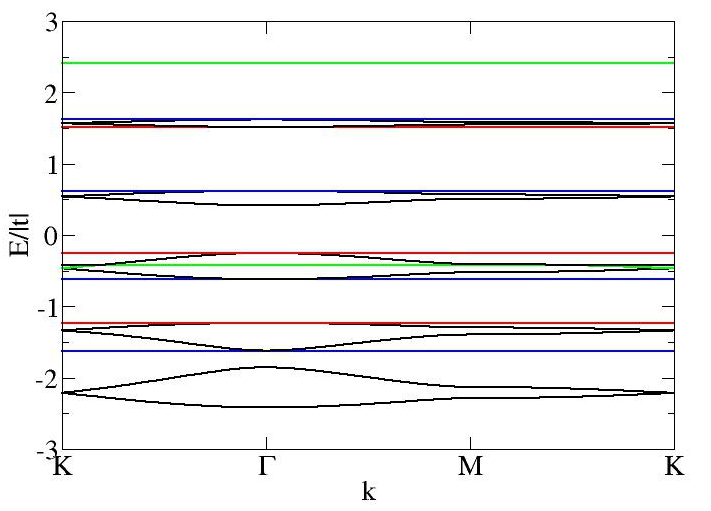} } \\
	\caption{\label{fig:decorated-honeycomb-kagome-band} Band structure of the ligand decorated honeycomb-Kagome lattice for various values of $t'/t$. top: $t'/t=0.5$, middle $t/t=1$ and bottom $t'/t=2$. In the latter case the three(two dispersive and one Kagome flat band)  lower and upper bands   are out  of the energy range of the figure $|E/t|>3$. The dispersive bands are represented in black, the Kagome flat bands in red, the  ligand  flat bands of A$_2$ symmetry (doubly degenerated) in green and the ligand flat bands of E symmetry in blue. }
\end{figure}

These arguments are only valid for $t'/t\ll 1$  since one can neglect the interaction between the states. When $t'$ gets larger the situation becomes much more complex but for $t'$ sufficiently large, the band structure converges towards a fixed solution where the two lowest and highest dispersive bands and their Kagome flat bands are going to $\pm \infty$ and become flat. The other bands reach a stable configuration.
In Fig. \ref{fig:decorated-honeycomb-kagome-band} we show the  band structure of the ligand-decorated honeycomb kagome lattice for various values of $t'/t$. For $t'/t=0.5$  it is clear that the bands near the the zero of energy are the most dispersive. For $t'/t=1$ the amplitude of the dispersive bands increases  and  shifts upward or downward such that some of them are  in contact and blocked by a E flat band. For $t'/t=2$ the band structure has almost reached a stable configuration. Note also that each time that dispersive bands are tangent to two flat bands one of them necessarily originates from E states. Finally one can mention that another type of flat band  can also exist in these MOFs when the metal is a transition metal dominated by $d$ electrons: if in a region of energy the spectrum of the ligand is dominated by $p_z$ orbitals (like in graphene around Fermi level) since $d_{z^2}$ does not couple to $p_z$ this will lead to three flat bands of $d_{z^2}$ character localized on the metallic atom.

\section{Conclusions}
\label{conclusions}
In summary, based on model tight-binding Hamiltonians we have derived the band structure of general classes of honeycomb-Kagome structures that can occur in various contexts, with a particular focus on MOFs. The energy spectrum of these systems is composed of dispersive bands similar to those of graphene with Dirac cones (and expressed in terms of the phase factor $\gamma(\bm{k})$) and four types of flat bands of very different nature. Some flat bands originates from states localized on a single entity (ligand or metal) which do not hybridize with other states for symmetry reasons, while other flat bands are built from correlated localized states forming hexagonal rings. A very interesting point in the band structure of these systems is their change of morphology with the parameters of the Hamiltonian (on-site and hopping integrals) such that one can easily modify the nature of these systems (semi-conductor, metallic, magnetic) by tuning the TB parameters. This opens up a wide range of possibilities to design a variety of materials and devices.
Note finally, that the analysis presented in this paper is a first step towards the understanding of 2D MOF electronic structure but further work is still  necessary to get a detailed description of realistic MOF, in particular the role of the various orbitals ($s$, $p$, $d$) of the atoms composing the material  has to be investigated thoroughly. However we believe that the methods developed in this work form a strong framework for further investigations of realistic materials. A forthcoming paper presenting DFT results of realistic MOFs is in preparation.

       \acknowledgments
The research leading to these results has received funding from the European Union H2020 Programme under grant agreement no. 696656 GrapheneCore1. C. Barreteau wishes to thank M. Brandbyge for its help in the use of Mathematica.

\clearpage

\end{document}